\documentclass[letterpaper]{IEEEtran}
\IEEEoverridecommandlockouts
\usepackage{cite}
\usepackage{amsmath,amssymb,amsfonts}
\usepackage{graphicx}
\usepackage{textcomp}
\usepackage{subcaption}
\usepackage{acronym}
\usepackage{romannum}
\usepackage{comment}
\usepackage{multirow}
\usepackage{algorithm}
\usepackage{algpseudocode}
\usepackage[table,xcdraw]{xcolor}
\usepackage{tikz}
\usetikzlibrary{
	positioning,
	shapes.geometric,
	decorations.pathreplacing,
	shadows,
	patterns,
	math,
	arrows.meta,
	shapes.multipart,
	calc,
    fit
}
\usepackage{circuitikz}
\usepackage{minted}
\usepackage{listings}

\definecolor{lightgray}{rgb}{0.95, 0.95, 0.95}
\definecolor{darkgray176}{RGB}{176,176,176}
\definecolor{darkorange25512714}{RGB}{255,127,14}
\definecolor{forestgreen4416044}{RGB}{44,160,44}
\definecolor{lightgray204}{RGB}{204,204,204}
\definecolor{steelblue31119180}{RGB}{31,119,180}

\definecolor{dur_color_light}{HTML}{E8A673}
\definecolor{dur_color_regular}{HTML}{D55E00}
\definecolor{dur_color_dark}{HTML}{8A3D00}

\definecolor{urb_color_light}{HTML}{73B1D5}
\definecolor{urb_color_regular}{HTML}{0072B2}
\definecolor{urb_color_dark}{HTML}{004A74}

\definecolor{sur_color_light}{HTML}{73CAB2}
\definecolor{sur_color_regular}{HTML}{009E73}
\definecolor{sur_color_dark}{HTML}{00674B}

\definecolor{all_color_light}{HTML}{E3B5CF}
\definecolor{all_color_regular}{HTML}{CC79A7}
\definecolor{all_color_dark}{HTML}{854F6D}

\definecolor{traditional_color_light}{HTML}{B8B8B8}
\definecolor{traditional_color_dark}{HTML}{3A3A3A}

\definecolor{selector_color_light}{HTML}{5FD6EA}
\definecolor{selector_color_dark}{HTML}{00809E}

\definecolor{combiner_color_light}{HTML}{00fd00}
\definecolor{combiner_color_dark}{HTML}{00b000}

\definecolor{signals_and_data_color}{HTML}{FFFFFF}
\definecolor{fixed_processing_color}{HTML}{E6E9EE}
\definecolor{channel_color}{HTML}{C9CFD8}
\definecolor{learned_layers_color}{HTML}{DDE5A8}
\definecolor{group_box_color}{HTML}{F3F5E3}

\usepackage{pgfplots}
\pgfplotsset{compat=1.16}

\usepackage{breqn}
\usepackage{siunitx}
\newacro{NTN}{Non-Terrestrial Network}
\newacroplural{NTN}[NTNs]{Non-Terrestrial Networks}
\newacro{TN}{Terrestrial Network}
\newacroplural{TN}[TNs]{Terrestrial Networks}
\newacro{LSP}{Large Scale Parameter}
\newacro{LS}{Least Squares}
\newacroplural{LSP}[LSPs]{Large Scale Parameters}
\newacro{SSP}{Small Scale Parameter}
\newacroplural{SSP}[SSPs]{Small Scale Parameters}
\newacro{ML}{Machine Learning}
\newacro{LoS}{Line-of-Sight}
\newacro{NLoS}{Non-Line-of-Sight}
\newacro{UL}{uplink}
\newacro{DL}{downlink}
\newacro{GPU}{Graphics Processing Unit}
\newacro{3GPP}{3rd Generation Partnership Project}
\newacro{FSPL}{Free-Space Path Loss}
\newacro{CIR}{Channel Impulse Response}
\newacroplural{CIR}[CIRs]{Channel Impulse Responses}
\newacro{BS}{Base Station}
\newacroplural{BS}[BSs]{Base Stations}
\newacro{OFDM}{Orthogonal Frequency-Division Multiplexing}
\newacro{SNR}{Signal to Noise Ratio}
\newacro{ITU}{International Telecommunication Union}
\newacro{NN}{Neural Network}
\newacro{BER}{Bit Error Rate}
\newacro{BLER}{Block Error Rate}
\newacro{CSI}{Channel State Information}
\newacro{SISO}{Single Input Single Output}
\newacro{MISO}{Multiple Input Single Output}
\newacro{SIMO}{Single Input Multiple Output}
\newacro{LDPC}{Low Density Parity Check}
\newacro{LMMSE}{Linear Minimum Mean Squared Error}
\newacro{MIMO}{Multiple Input Multiple Output}
\newacro{GEO}{Geostationary Orbit}
\newacro{LEO}{Low Earth Orbit}
\newacro{MEO}{Medium Earth Orbit}
\newacro{HAPS}{High Altitude Platform Station}
\newacro{UE}{User Equipment}
\newacro{VSAT}{Very Small Aperture Terminal}
\newacro{DS}{delay spread}
\newacro{ZOA}{Zenith of Arrival}
\newacro{ZOD}{Zenith of Departure}
\newacro{UT}{User Terminal}
\newacroplural{UT}[UTs]{User Terminals}
\newacro{TDL}{Tapped Delay Line}
\newacro{CDL}{Cluster Delay Line}
\newacro{CP}{Cyclic Prefix}
\newacro{ISI}{Inter Symbol Interference}
\newacro{SF}{Shadow Fading}
\newacro{AOD}{Angle of Departure}
\newacro{AOA}{Angle of Arrival}
\newacro{KPI}{Key Performance Indicator}
\newacroplural{KPI}[KPIs]{Key Performance Indicators}
\newacro{LLR}{Log Likelihood Ratio}
\newacroplural{LLR}[LLRs]{Log Likelihood Ratios}
\newacro{HPDPU}{High Performance Data Processing Unit}
\newacro{URB}{Urban}
\newacro{DUR}{Dense Urban}
\newacro{SUR}{Suburban}
\newacro{MoE}{Mixture-of-Experts}
\newacro{SCE}{Sigmoid Cross-Entropy}
\newacro{KD}{Knowledge Distillation}

\def\BibTeX{{\rm B\kern-.05em{\sc i\kern-.025em b}\kern-.08em
    T\kern-.1667em\lower.7ex\hbox{E}\kern-.125emX}}

\begin{document}

\title{Towards Deployable Learned Receivers for Non-Terrestrial Networks\\

\thanks{This research was supported in part by the German Federal Ministry of Research, Technology and Space (BMFTR) within the projects Open6GHub+ under grant number 16KIS2408 and 6G-Coverage under grant number 16KIS2428K as well as the European Space Agency (ESA) under contract number 4000139559/22/UK/AL (AIComS).}
}

\author{\IEEEauthorblockN{Tim Düe, Fayad Haddad, Dirk Wübben, Carsten Bockelmann, Armin Dekorsy\\}
\IEEEauthorblockA{\textit{Department of Communications Engineering},
\textit{University of Bremen},
Bremen, Germany. \\
\{duee, haddad, wuebben, bockelmann, dekorsy\}@ant.uni-bremen.de}\\}

\maketitle
\thispagestyle{empty}

\begin{abstract}
\acp{NTN} are expected to become integral to future 6G systems, but their heterogeneous, environment-dependent channel conditions challenge receivers, while satellite platforms impose far tighter compute, energy, and memory budgets than terrestrial platforms. Learned physical-layer receivers have shown strong gains in terrestrial systems, but their deployability on satellite hardware remains largely unexplored. This work investigates this using a learned receiver inspired by the DeepRx architecture, trained and evaluated under standardized 3GPP TR 38.811 NTN channel models. The results reveal a cross-scenario mismatch, whereby training and evaluating on the same channel does not necessarily yield the best performance, which is particularly important to satellites that need to operate on heterogeneous channels simultaneously. To address this challenge, \ac{MoE} architectures incorporating multiple scenario-specific experts are investigated. Furthermore, the performance-complexity trade-off is characterized through scaling, pruning, and knowledge distillation. Finally, the receiver is synthesized for an existing satellite ML hardware platform, outlining a pathway toward deployable NTN receivers.
\end{abstract}


\section{Introduction}

\acfp{NTN} are a key technology for 6G, extending coverage to remote areas, oceans, the air, and disaster regions, with satellites playing a central role due to their constant presence and large-area coverage~\cite{3gpp.38.821}. However, NTN propagation conditions vary strongly with the user environment and satellite elevation angle~\cite{3gpp.38.811, 3gpp.38.821}, and satellites must handle this heterogeneity simultaneously as they fly over different regions serving users with varying channel characteristics, all under tightly constrained on-board computational resources.

In \acp{TN}, \ac{ML}-based communications has shown strong results, with a particularly popular category being learned neural receivers based on the DeepRx architecture introduced in~\cite{deeprx}. This architecture jointly performs channel estimation, equalization, and demapping in a single learned block. Since its original introduction as a \ac{SIMO} \ac{OFDM} receiver, it has been extended for several use cases~\cite{MIMO_DeepRx, HybridDeepRX, neural_receiver} and shown to be real-time capable on actual hardware~\cite{wiesmayr2025}.

Neural receivers have likewise been proposed for satellite communications~\cite{zhang2020leo, nielsen2022}, including work that applies the same core architectural element as DeepRx, the residual block~\cite{resnet}, to channel estimation~\cite{abdollahpour2026}. Residual-block architectures have further been investigated with regard to compute- and memory-efficient designs, though without an explicit satellite focus~\cite{abdollahpour2025}. This work investigates the use of DeepRx-inspired neural receivers for the satellite use case.

Our investigations identify a gap in the literature: learned receivers are frequently claimed to generalize inherently, i.e., a receiver trained on one channel continues to perform well on a different one~\cite{deeprx, neural_receiver, abdollahpour2026}, but, to our knowledge, this claim has not been systematically evaluated across scenarios. This is particularly relevant for satellites, which, unlike stationary terrestrial base stations, fly over different locations and must serve users under different channel conditions. The results show that matching training and evaluation conditions does not always yield the best performance, a mismatch that, to our knowledge, has not been previously reported. To address this issue, \acf{MoE}-based solutions are proposed. Finally, the model is synthesized for an existing satellite hardware platform, and its deployment characteristics are reported.

The main contributions are:
\begin{itemize}
    \item An evaluation of DeepRx-inspired neural receivers on realistic 3GPP~TR~38.811 NTN channels.
    \item \ac{MoE}-based architectures for improving performance under heterogeneous channel conditions.
    \item A characterization of the performance-complexity trade-off of the neural receiver.
    \item Synthesis of the neural receiver for a hardware accelerator platform, including an evaluation of its performance metrics.
\end{itemize}

\vspace{-11pt}
\section{System Model}

A \ac{SIMO} \ac{OFDM} \ac{UL} transmission is considered. Information bits are encoded with a \ac{LDPC} code of rate $R$ and mapped to a 4QAM constellation $\mathcal{C}$. The \ac{OFDM} grid consists of $N_{\text{FFT}}$ subcarriers and $N_{\text{Sym}}$ symbols. The satellite receiver is equipped with $N_{\text{Ant}}$ antennas, while the ground-based transmitter uses a single antenna. Ideal Doppler and timing synchronization are assumed, as achievable using the methods described in~\cite{3gpp_ts38331, 3gpp.38.821}, since uncompensated Doppler shifts can destroy subcarrier orthogonality in \acp{NTN}.

The \ac{NTN} channel is a time- and frequency-selective multipath channel described by its \ac{CIR}. At each time instant $t$, the channel between the transmitter and receive antenna $a$ is characterized by $P$ paths, with time-varying path amplitudes $\alpha_a(t) \in \mathbb{C}^{P}$ and delays $\tau_a \in \mathbb{R}^{P}$. The effective channel coefficient is obtained by superpositioning the paths,
\begin{equation}
    h_a(t,f) = \sum_{p=1}^P \alpha_{a,p}(t) \cdot e^{-j 2 \pi f \tau_{a,p}}.
\end{equation}

For each \ac{OFDM} resource element at time $t$ and subcarrier $f$ the signal received at antenna $a$ is
\begin{equation}
    y_a(t,f) = h_a(t,f) \cdot x(t,f) + w_a(t,f),
\end{equation}
where $x(t,f) \in \mathcal{C}$ is the transmitted symbol, $h_a(t,f) \in \mathbb{C}$ the effective channel coefficient between transmitter and receive antenna $a$, and $w_a(t,f) \in \mathbb{C}$ additive white Gaussian noise with spectral noise density $N_0 \in \mathbb{R}$. Stacking all resource elements and antennas yields the received grid $\mathbf{Y} \in \mathbb{C}^{N_{\text{Sym}} \times N_{\text{FFT}} \times N_{\text{Ant}}}$, which, together with $N_0$, forms the input to both the conventional and learned receivers.

Conventional signal processing is done using the same approach found in literature \cite{6GNTN, deeprx}. The channel estimate $\tilde h_a(t,f)$ is derived using a \ac{LS} estimator on pilot symbols, interpolated linearly over time. Stacking the per-antenna estimates for a given resource element gives $\tilde{\mathbf{h}}(t,f) \in \mathbb{C}^{N_{\text{Ant}}}$, and equalization is performed per resource element via a \ac{LMMSE} equalizer,
\begin{equation}\label{eq:lmmse}
\begin{aligned}
    \mathbf{g}(t,f) &= \Big( \tilde{\mathbf{h}}(t,f)\tilde{\mathbf{h}}^H(t,f) + N_0\mathbf{I} \Big)^{-1} \tilde{\mathbf{h}}(t,f), \quad \\
    \tilde{x}(t,f) &= \mathbf{g}^H(t,f) \mathbf{y}(t,f),
\end{aligned}
\end{equation}
where $\mathbf{I}$ is the identity matrix, $(\cdot)^H$ the Hermitian transpose, and $\mathbf{y}(t,f) \in \mathbb{C}^{N_{\text{Ant}}}$ the received vector across all antennas for this resource element. Each \ac{OFDM} frame consists of the code bits ${\mathbf{C}} \in \{0,1\}^{N_{\text{Sym}} \times N_{\text{FFT}} \times N_{\text{B}}}$, where $N_{\text{B}}$ refers to the number of bits per symbol. Consistent with \cite{wiesmayr2025}, the \ac{LLR} $\tilde{\mathbf{L}} \in \mathbb{R}^{N_{\text{Sym}} \times N_{\text{FFT}} \times N_{\text{B}}}$ are calculated as
\begin{equation}
    \tilde{l}(t,f,i) = \ln\frac{\sum_{x\in \mathcal{C}_{i,1}} \exp(-|\tilde{x}(t,f)-x|^2/N_{\text{0}})}{\sum_{x\in \mathcal{C}_{i,0}} \exp(-|\tilde{x}(t,f)-x|^2/N_{\text{0}})},
\end{equation}
where $\mathcal{C}_{i,0}$ and $\mathcal{C}_{i,1}$ mark all symbols of constellation $\mathcal{C}$ for which bit $i$ is 0 or 1 respectively.

\vspace{-10pt}

\section{Model Architectures and Training}\label{sec:architecture_and_training}

Satellites need to operate under various channel scenarios. This section first introduces the structure and training of a neural receiver optimized in a single channel scenario, followed by our \ac{MoE} architectures combining multiple receivers aiming to improve the performance when applied across scenarios.

\vspace{-10pt}

\subsection{Channel Scenario Specific Neural Receivers}

\begin{figure}[htbp]
    \centering
    \resizebox{0.9\columnwidth}{!}{\begin{tikzpicture}[
    >=Latex,
    node distance=0.9cm,
    font=\Large,
    block/.style={
        draw,
        rounded corners=4pt,
        minimum width=3.2cm,
        minimum height=1cm,
        align=center,
        line width=1.5pt
    },
    green/.style={block,draw=green!97!black!31!gray,fill=green!80!black!10,},
    blue/.style={block,draw=blue!50!black,fill=blue!20},
    purple/.style={block,draw=violet!60!black,fill=violet!20},
    yellow/.style={block,draw=orange!70!black,fill=yellow!25},
    red/.style={block,draw=red!70!black,fill=red!20},
    oval/.style={
        draw=violet!60!black,
        fill=violet!20,
        ellipse,
        minimum width=2.2cm,
        minimum height=1cm,
        line width=1.5pt
    },
    flow/.style={-Latex,very thick},
    dashbox/.style={
        draw=blue!40,
        dashed,
        line width=1.5pt
    },
]

\draw [decorate, decoration={brace,amplitude=10pt,mirror,raise=4ex}, ultra thick]
  (-1.71,-5.19) -- (-1.71,-7.19) node[yshift=-2.5em, xshift=-5.5em, minimum width=35pt, at start]{\LARGE$N_{\text{Blocks}}$};


\node[green, draw=black, fill=white] (input) {$\mathbf{Y}$};

\node[blue, below=0.8cm of input, fill=fixed_processing_color, draw=black] (seperateDim) {Concat [$\Re$,$\Im$,$N_0$]};
\node[green, below=0.8cm of seperateDim] (preprocessing) {Input Conv};
\node[purple,below=0.8cm of preprocessing] (r1) {Residual Block};
\node[purple, below=0.3cm of r1] (r2) {Residual Block};

\node[green, below=0.8cm of r2] (outputConv) {Output Conv};

\node[green, below=0.8cm of outputConv, draw=black, fill=white] (out) {$\tilde{\mathbf{L}}$};

\draw[flow] (input)--(seperateDim);
\draw[flow] (seperateDim)--(preprocessing);
\draw[flow] (preprocessing)--(r1);
\draw[flow] (r1)--(r2);
\draw[flow] (r2)--(outputConv);
\draw[flow] (outputConv)--(out);

\node[purple] (r3) at (-0.15,-7.05) {Residual Block};
\node[purple] (r4) at (-0.3,-7.2) {Residual Block};


\node[circle,right=3.5cm of input] (a0) {};

\node[purple, draw=black, fill=fixed_processing_color, below right=0.8cm and 0.7cm of a0] (n1)
{Norm};

\node[yellow, draw=black, fill=fixed_processing_color, below=0.55cm of n1]
(relu1){ReLU};

\node[green,below=0.55cm of relu1]
(conv1){Conv2D};

\node[purple, draw=black, fill=fixed_processing_color, below=0.55cm of conv1]
(n2){Norm};

\node[yellow, draw=black, fill=fixed_processing_color, below=0.55cm of n2]
(relu2){ReLU};

\node[green,below=0.55cm of relu2]
(conv2){Conv2D};

\node[circle, draw=black, fill=red!0, line width=1.5pt, minimum size=8mm, inner sep=0pt, font=\LARGE\bfseries] (additionSkip)
at ($(a0.south |- conv2.west)$)
{+};

\node[circle,below=0.9cm of additionSkip] (a1) {};



\draw[flow] (n1)--(relu1);
\draw[flow] (relu1)--(conv1);
\draw[flow] (conv1)--(n2);
\draw[flow] (n2)--(relu2);
\draw[flow] (relu2)--(conv2);



\node[dashbox, fit=(a0) (n1) (relu1) (conv1) (n2) (relu2) (conv2), inner sep=0.2cm] (dashLarge) {};

\node[dashbox,
fit=(r1),
inner sep=0.1cm] (dashSmall){};

\draw[->, very thick] (5.3,0) -| (n1.north);

\draw[flow] (5.3,0.75) -- (additionSkip);
\draw[flow] (conv2) -- (additionSkip);
\draw[flow] (additionSkip) -- (a1);

  \draw[blue!40, dashed, ultra thick] (dashLarge.north west) -- (dashSmall.north east);
  \draw[blue!40, dashed, ultra thick] (dashLarge.south west) -- (dashSmall.south east);
  \node[] at (-3.53,-0.93) {\LARGE$\mathbb{C}^{N_{\text{Sym}} \times N_{\text{FFT}} \times N_{\text{Ant}}}$};
\node[] at (-3.46,-2.83) {\LARGE$\mathbb{R}^{N_{\text{Sym}} \times N_{\text{FFT}} \times  2 \cdot N_{\text{Ant}} + 1}$};
\node[] at (-3.50,-4.6) {\LARGE$\mathbb{R}^{N_{\text{Sym}} \times N_{\text{FFT}} \times N_{\text{Conv}}}$};
\node[] at (-3.50,-8.04) {\LARGE$\mathbb{R}^{N_{\text{Sym}} \times N_{\text{FFT}} \times N_{\text{Conv}}}$};
\node[] at (-3.5,-9.79) {\LARGE$\mathbb{R}^{N_{\text{Sym}} \times N_{\text{FFT}} \times N_{\text{B}}}$};

\end{tikzpicture}}
    \caption{Structure of the residual-block based learned receiver.}
    \label{fig:drx_structure}
\end{figure}
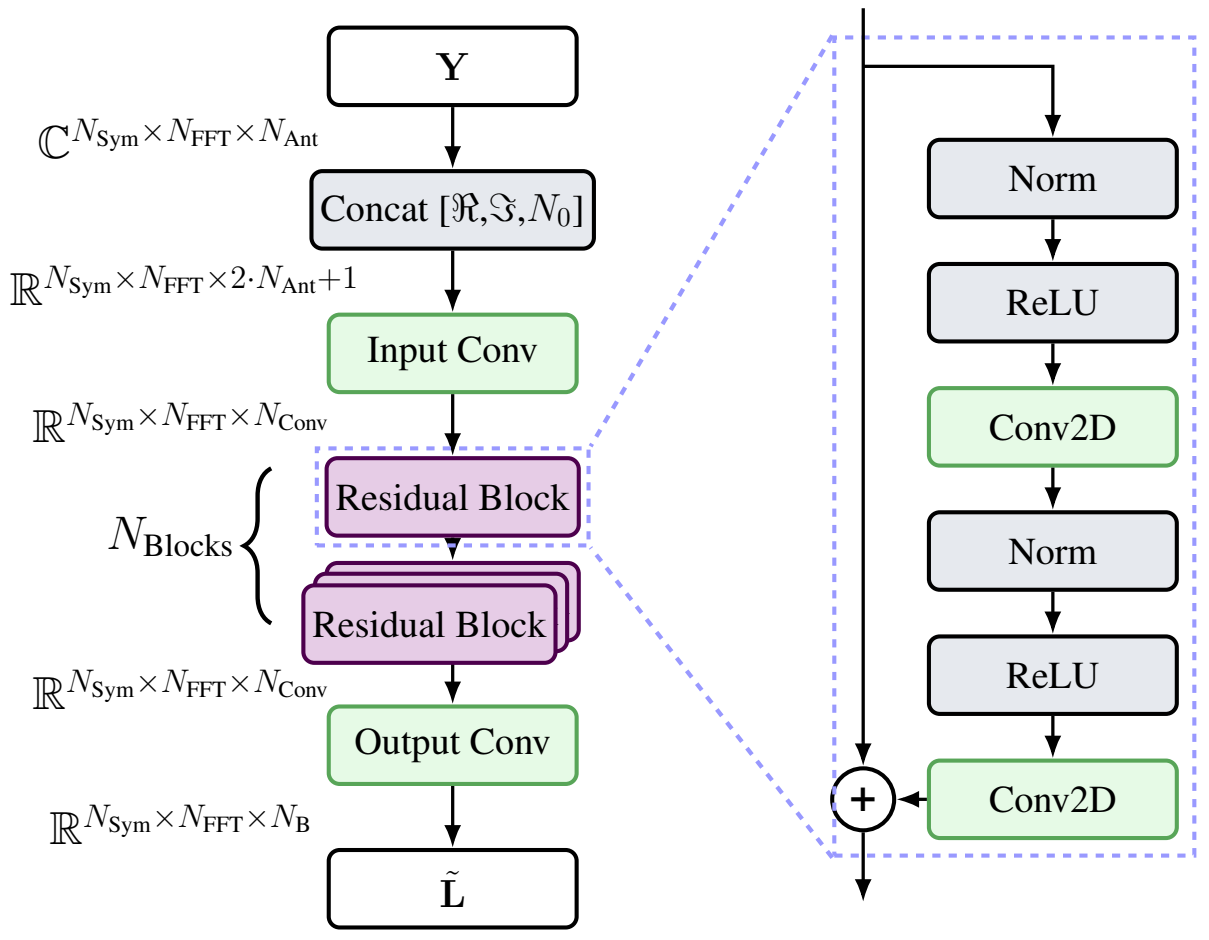

The neural receiver $\mathcal{M}$ is a fully convolutional \ac{ML} model built from residual blocks inspired by the DeepRx architecture~\cite{deeprx}, mapping the signal $\mathbf{Y}$ to bit reliability estimates $\tilde{\mathbf{L}}$. As shown in Fig.~\ref{fig:drx_structure}, the real and imaginary parts of $\mathbf{Y}$ and $N_0$ are concatenated (Concat) and mapped to $N_{\text{Conv}}$ convolutional channels via an input convolution. The term convolutional channels in the context of \ac{ML} refers to the number of extracted features for a convolutional layer and is not to be confused with a communications channel. A sequence of residual blocks then successively applies normalization, ReLU, and convolutions with a skip connection, before a final convolution produces $\tilde{\mathbf{L}}$. The main hyperparameters of $\mathcal{M}$ are its depth (number of residual blocks $N_{\text{Blocks}}$) and width (number of convolutional channels $N_{\text{Conv}}$).
Applying $\mathcal{M}$ is written as
\begin{equation}\label{eq:model_function}
    \tilde{\mathbf{{L}}}^k = \mathcal{M}_{\boldsymbol{\theta}}^k(\mathbf{Y},N_0),
\end{equation}
where $\boldsymbol{\theta}$ is the parametrization of the model. The superscript $k$ is an additional name to differentiate between different models.

The objective when training $\mathcal{M}^k$ is to find the optimal parameterization $\boldsymbol{\theta}^*$ that minimizes the expected \ac{SCE} between the predicted LLRs $\tilde{\mathbf{L}}$ and the corresponding code bits $\mathbf{C}$. For a set of $N$ training samples, the objective is expressed as
\begin{equation}
\label{eq:sce_loss}
\begin{aligned}
\mathcal{L}_{\text{SCE}}(\mathbf{C},\tilde{\mathbf{L}}) = -\frac{1}{N}\sum_{t,f,i}
&\Big[  c_{t,f,i}\log\sigma(\tilde{l}_{t,f,i}) \\
 &+ (1-c_{t,f,i})\log\big(1-\sigma(\tilde{l}_{t,f,i})\big) \Big],
\end{aligned}
\end{equation}
where $\sigma$ is the sigmoid function. The formal training objective
\begin{equation}
    \boldsymbol{\theta}^{*} = \operatorname*{argmin}_{\boldsymbol{\theta}} \mathbb{E}_{(\mathbf{C},\mathbf{Y},N_0) \sim \mathcal{D}}\Big\{\mathcal{L}_{\text{SCE}}\Big( \mathbf{C}, \mathcal{M}_{\boldsymbol{\theta}}^k(\mathbf{Y},N_0)\Big)\Big\}
\end{equation}
of minimizing the loss over the training dataset $\mathcal{D}$ has been used for these architectures since their introduction~\cite{deeprx} and, to our knowledge, remains the state-of-the-art across extensions~\cite{HybridDeepRX,MIMO_DeepRx,neural_receiver,nielsen2022,abdollahpour2025}.

\vspace{-10pt}

\subsection{Mixture-of-Experts Based Architectures}

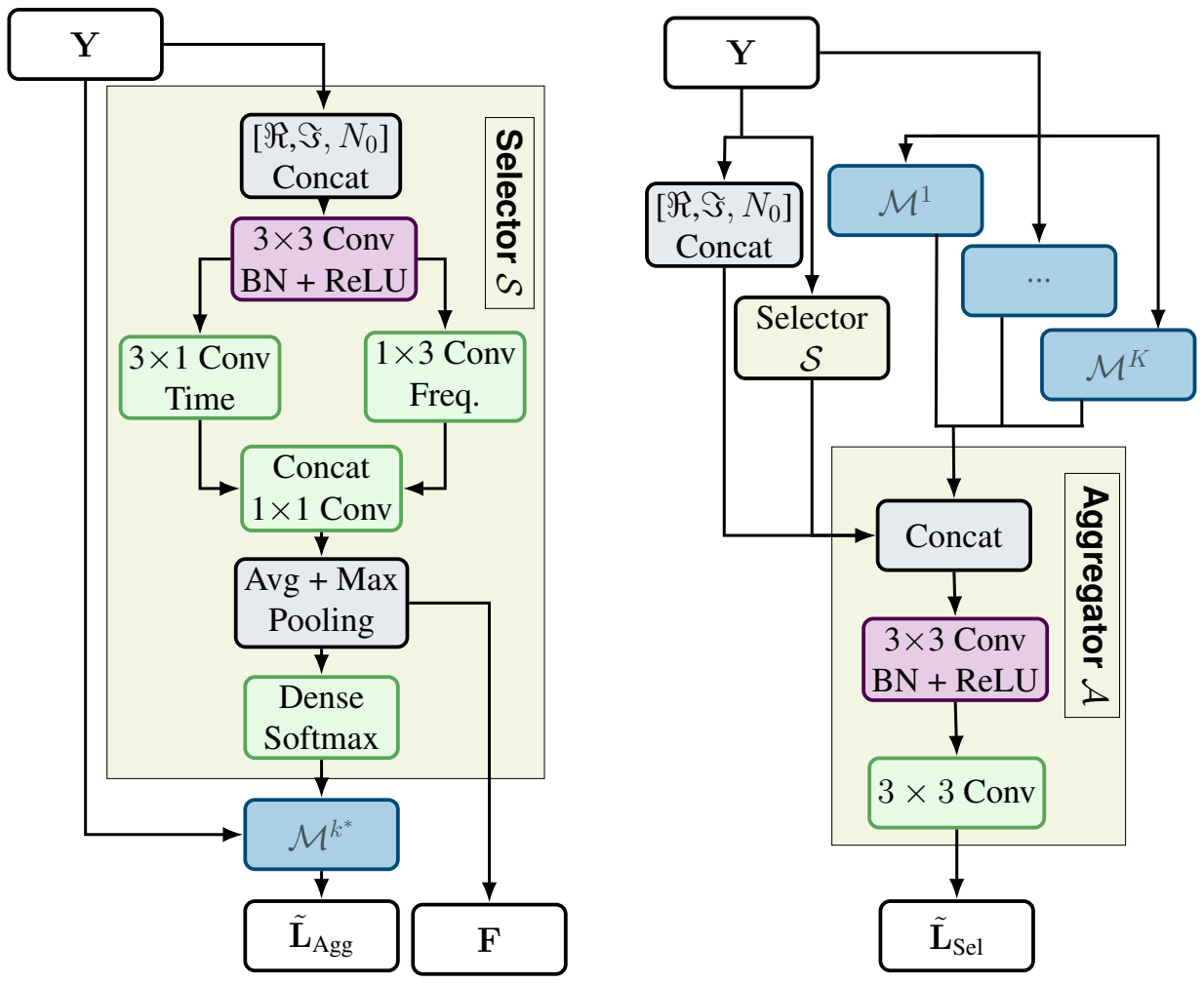
\begin{figure}
    \hspace{10pt}
    \resizebox{1.0\columnwidth}{!}{\begin{tikzpicture}[
    >=Latex,
    font=\Large,
    block/.style={
        draw,
        rounded corners=4pt,
        minimum width=2.2cm,
        minimum height=1cm,
        align=center,
        line width=1.5pt
    },
    green/.style={block,draw=green!97!black!31!gray,fill=green!80!black!10,},
    blue/.style={block,draw=blue!50!black,fill=blue!20},
    purple/.style={block,draw=violet!60!black,fill=violet!20},
    yellow/.style={block,draw=orange!70!black,fill=yellow!25},
    red/.style={block,draw=red!70!black,fill=red!20},
    oval/.style={
        draw=violet!60!black,
        fill=violet!20,
        ellipse,
        minimum width=2.2cm,
        minimum height=1cm,
        line width=1.5pt
    },
    flow/.style={-Latex,very thick},
    dashbox/.style={
        draw=blue!40,
        dashed,
        line width=1.5pt
    },
]
\definecolor{signals_and_data_color}{HTML}{FFFFFF}
\definecolor{fixed_processing_color}{HTML}{E6E9EE}
\definecolor{channel_color}{HTML}{C9CFD8}
\definecolor{learned_layers_color}{HTML}{DDE5A8}
\definecolor{group_box_color}{HTML}{F3F5E3}
\definecolor{dur_color_light}{HTML}{E8A673}
\definecolor{dur_color_regular}{HTML}{D55E00}
\definecolor{dur_color_dark}{HTML}{8A3D00}
\definecolor{urb_color_light}{HTML}{73B1D5}
\definecolor{urb_color_regular}{HTML}{0072B2}
\definecolor{urb_color_dark}{HTML}{004A74}
\definecolor{sur_color_light}{HTML}{73CAB2}
\definecolor{sur_color_regular}{HTML}{009E73}
\definecolor{sur_color_dark}{HTML}{00674B}
\draw[fill=group_box_color] (-17.14,0.9) rectangle (-10.84,-9.08);
\draw[fill=group_box_color] (-6.7,-4.3) rectangle (-2.47,-10.04);
\node[blue, above left={1.44cm and 4.54cm}, fill=fixed_processing_color, draw=black, minimum width=65pt] (featureL2)
      at (-8.35,-2.17) {[$\Re$,$\Im,N_0$]\\Concat};
\node[green, draw=black, fill=white] (inputL)
    at (-8,1.38) {${\mathbf{Y}}$};
\node[blue,
      below right={0.78cm and 2.82cm},
      fill=urb_color_light,
      draw=urb_color_dark,
      fill opacity=0.6] (urbL)
    at (-7.66,-0.6) {...};
\node[blue,
      below right={0.27cm and 1.96cm},
      fill=urb_color_light,
      draw=urb_color_dark,
      fill opacity=0.6] (durL)
    at (-5.66,-2.32) {\Large$\mathcal{M}^{K}$};
\node[blue,
      left=8.02cm,
      fill=urb_color_light, draw=urb_color_dark, fill opacity=0.6] (surL)
    at (3.51,-0.75) {\Large$\mathcal{M}^1$};
\node[blue, left=1.78cm, fill=fixed_processing_color, draw=black] (featureL)
    at (-5.33,-1.1) {[$\Re$,$\Im,N_0$]\\Concat};
\node[blue,
      left=4.66cm,
      fill=group_box_color,
      draw=black] (selectorL)
    at (-1.20,-2.73) {Selector\\$\mathcal{S}$};
\node[blue,
      below=1.99cm,
      fill=fixed_processing_color,
      draw=black] (concatL)
    at (-4.94,-3.06) {Concat};
\node[blue,
      below=0.25cm,
      draw=green!97!black!31!gray, 
      fill=green!80!black!10, ] (conv2L)
    at (-4.90,-8.51) {$3\times3$ Conv};
\node[blue,
      below=1cm,
      fill=white,
      draw=black] (llrL)
    at (-4.9,-9.82) {$\tilde{\mathbf{L}}_{\text{Sel}}$};
\draw[flow]
    (inputL) -| (urbL);
\draw[flow]
    (-4.84,0.18) -| (surL.north);
\draw[flow]
    (-5.24,0.18) -| (-2,-2.62);
\draw[flow]
    (-8.24,0.18) -- (featureL.north);
\draw[flow]
    (-8.24,0.16) -| (selectorL.north);
\draw[flow]
    (-4.95,-4) -- (concatL.north);
\draw[flow]
    (conv2L) -- (llrL);
\node[green, draw=black, fill=white]
    (inputR) at (-17.45,1.5)
    {${\mathbf{Y}}$};
\node[purple,
      draw=violet!60!black,
      fill=violet!20,
      below right={1.93cm and 0.75cm}] (featureL2)
    at (-16.11,0.95) {3$\times$3 Conv\\BN + ReLU};
\node[
      minimum size=0.0cm
] (split) {};
\node[blue, 
      draw=green!97!black!31!gray, 
      fill=green!80!black!10, 
      ] (time)
    at (-15.8,-3.3) {3$\times$1 Conv\\Time};
\node[blue,
      draw=green!97!black!31!gray,        fill=green!80!black!10, 
      ] (freq)
    at (-12.27,-3.29) {1$\times$3 Conv\\Freq.};
\node[blue,
      draw=green!97!black!31!gray,        fill=green!80!black!10, 
    ] (fusion)
    at (-14.05,-4.9) {Concat\\1$\times$1 Conv};
\node[blue,
      draw=black,
      fill=fixed_processing_color,
      below=0.24cm] (pool)
    at (-14.05,-5.65) {Avg + Max\\Pooling};
\node[blue,
      draw=green!97!black!31!gray,        fill=green!80!black!10, 
      below=0.29cm] (sel7)
    at (-14.05,-7.3) {Dense\\Softmax};
\node[blue,
      below=0.51cm,
      fill=urb_color_light,
      draw=urb_color_dark,
      fill opacity=0.6] (durR)
    at (-14.05,-8.85) {\Large$\mathcal{M}^{{k^*}}$};
\node[blue,
      below=1.17cm,
      fill=white,
      draw=black] (llrR)
    at (-14.05,-9.65) {$\tilde{\mathbf{L}}_{\text{Agg}}$};
\node[blue,
      below right={1.17cm and 1.29cm},
      fill=white,
      draw=black] (featuresR)
    at (-14.05,-9.7) {${\mathbf{F}}$};
\draw[flow]
    (time) |- (fusion);
\draw[flow]
    (freq) |- (fusion);
\draw[flow]
    (fusion) -- (pool);
\draw[flow]
    (pool) -- (sel7);
\draw[flow]
    (pool.east) -| (featuresR.north);
\draw[flow]
    (inputR.south) |- (durR.west);
\node[draw, node font=\Large\sffamily\bfseries, fill=group_box_color, rotate=-90] at (-11.34,-0.93) {Selector $\mathcal{S}$};
\node[draw, node font=\Large\sffamily\bfseries, fill=group_box_color, rotate=-90] at (-2.95,-6.44) {Aggregator $\mathcal{A}$};
\draw[flow] (inputR.east) -| (-14.02,0.54);
\draw[flow] (-14.02,-0.7) -- (featureL2);
\draw[flow] (sel7.south) -- (durR.north);
\draw[flow] (durR.south) -- (llrR.north);
\draw[flow] (featureL2.east) -| (freq.north);
\draw[flow] (featureL2.west) -| (time.north);
\node[purple,
        draw=violet!60!black,
        fill=violet!20,
        below right={7.06cm and 8.96cm}] (featureL3)
      at (-15.23,0.31) {3$\times$3 Conv\\BN + ReLU};
\draw[flow]
    (featureL.south) |- (concatL.west);
\draw[flow]
    (selectorL.south) |- (concatL.west);
\draw[flow]
    (concatL.south) -- (featureL3.north);
    
\draw[flow]
    (featureL3.south) -- (conv2L.north);
\draw[very thick] (-5.64,0.18) -- (-3.5,0.18);
\draw[very thick] (-5.22,-4) -- (-3.07,-4);
\draw[very thick] (-5.2,-4.02) -- (-5.19,-1.23);
\draw[very thick] (-4.25,-4) -- (-4.25,-2.43);
\draw[very thick] (-3.1,-4) -- (-3.1,-3.62);
\draw[very thick] (-8,0.9) -- (-8,0.18);
\end{tikzpicture}}
    \caption{Mixture-of-Experts architectures. \textit{Left}: selective architecture forwards signal to selected expert. \textit{Right}: aggregating architecture combines signal with selector and expert outputs.}
    \label{fig:multihead_architectures}
\end{figure}

$K$ dedicated neural receivers (experts), each for a specific scenario, are trained and combined using the architectures shown in Fig.~\ref{fig:multihead_architectures}.

The left side of Fig.~\ref{fig:multihead_architectures} shows the architecture of the selective model $\mathcal{M}^{\text{Sel}}$. It consists of a selector $\mathcal{S}$ and $K$ pretrained experts. $\mathcal{S}$ predicts which expert performs best for the current signal $\mathbf{Y}$ and forwards it accordingly.

To operate in a feature space relevant to the experts, $\mathcal{S}$ first mirrors their feature extraction with half a residual block, followed by parallel $3{\times}1$ and $1{\times}3$ convolutions fused by a $1{\times}1$ convolution to extract time and frequency features. Global Avg and Max pooling then extract average and outlier features. A final fully connected layer with softmax predicts for each expert the probability that it performs best for the input signal. The pooled features $\mathbf{F}$ are also returned for later use in the aggregating model.
\begin{equation}
S : \mathbf{Y} \mapsto \left(s^{{1}},\,...,\,s^{K}, \mathbf{F}\right), \quad
\textstyle\sum_{k=1}^{K} s^{k} = 1.
\end{equation}
During training the weights of all experts $\mathcal{M}^k$ remain frozen and only the selector $\mathcal{S}$ is trained as
\begin{equation}\label{eq:optimization_objective_selective}
    \boldsymbol{\Psi}^* = \operatorname*{argmin}_{\boldsymbol{\Psi}} \; \sum_{k=1}^{K}
    s_{\boldsymbol{\Psi}}^{k}(\mathbf{Y})\cdot\,
    \mathcal{L}_{\text{SCE}}\big[\mathbf{C}, \mathcal{M}^k(\mathbf{Y},N_0)\big],
\end{equation}
the weighted sum of the experts' losses, weighted by $\mathcal{S}$'s predicted probabilities. Rather than discrete labels, $\mathcal{S}$ thus learns the relative performance of each expert, which the aggregator later uses to consider each expert's reliability. At inference, $\mathcal{M}^{\text{Sel}}$ forwards the signal to the expert with the highest probability,
\begin{equation}
    \mathcal{M}^{\text{Sel}}(\mathbf{Y}) = \mathcal{M}^{k^*}(\mathbf{Y}), \quad k^* = \operatorname*{argmax}_{k} s_{\boldsymbol{\Psi}}^{k}(\mathbf{Y}).
\end{equation}

Additionally, the aggregating architecture shown on the right side of Fig.~\ref{fig:multihead_architectures} is proposed, which forwards $\mathbf{Y}$ to all experts and combines the outputs in a learned manner.

$\mathcal{M}^{\text{Agg}}$ consists of $K$ experts, the selector $\mathcal{S}$ of $\mathcal{M}^{\text{Sel}}$, and an aggregator $\mathcal{A}$, which concatenates the experts' outputs with $\mathcal{S}$'s probabilities, a reliability estimate per expert, its pooled features $\mathbf{F}$, and the preprocessed $\mathbf{Y}$ itself, which retains information such as subcarrier reliabilities that may be lost in $\mathbf{F}$. $\mathcal{A}$ also uses half a residual block to combine information across experts, followed by a combining convolution.
When optimizing $\mathcal{M}^{\text{Agg}}$, only $\mathcal{A}$ is trained and the experts remain frozen.
\begin{equation}\label{eq:optimization_objective_combiner}
\boldsymbol{\Phi}^* = \operatorname*{argmin}_{\boldsymbol{\Phi}} \mathcal{L}_{\text{SCE}}\big[C_{\boldsymbol{\Phi}}(\mathbf{\tilde{L}}^{1}, \ldots, \mathbf{\tilde{L}}^{K}, \mathbf{s}, \mathbf{F}, \mathbf{Y}, N_0), \mathbf{C}\big].
\end{equation}

Additionally, $\mathcal{M}^{\text{Oracle}}$ provides an upper bound on the performance of a perfect selector. Using the true information bits $\mathbf{B}$, it evaluates all $K$ experts after decoding and selects the one yielding the best result:
\begin{equation}
    \label{eq:oracle}
    \begin{split}
    \mathcal{M}^{\text{Oracle}}(\mathbf{Y},N_0) = \mathcal{M}^{k^*}(\mathbf{Y},N_0), \;\;\\
    k^* = \operatorname*{argmin}_{k\in[1,K]} d_{H}\Big(\mathcal{D}\big(\mathcal{M}^{k}(\mathbf{Y},N_0)\big), \mathbf{B}\Big),
    \end{split}
\end{equation}
where $\mathcal{D}(\cdot)$ is decoding and $d_{H}(\cdot,\cdot)$ the Hamming distance.

\vspace{-15pt}
\section{Simulation Results}
\label{sec:baseline_perf}

This section first introduces the simulation setup, followed by the results for individual channels first and all channels second. To provide compatibility, the channel and system setups are chosen to be comparable with examples from literature, mainly~\cite[Sec.~5.2]{6GNTN} and \cite{zhang2020leo,abdollahpour2026}. 
\vspace{-15pt}
\subsection{Simulation Setup}

All channel realizations are generated using OpenNTN~\cite{openntn}. All investigations use a code rate of $R=\frac{526}{1024}$, $N_{\text{FFT}}=128$ subcarriers, $N_{\text{Sym}}=12$ symbols, and $N_{\text{Ant}}=8$ antennas. The satellite is at a height of \SI{600}{km}, the carrier frequency is \SI{3.5}{GHz}, subcarrier spacing is \SI{15}{kHz}, cyclic prefix length is 10, and two pilots over all subcarriers at symbol index 2 and 11 are used. 

12 channels are defined. Following~\cite{3gpp.38.811}, users are in one of three scenarios, \ac{DUR}, \ac{SUR}, or \ac{URB}, each defining characteristics such as mobility, losses, and \ac{LoS} probability. A fourth scenario, ALL, samples the scenario uniformly per transmission. Within each scenario, three \ac{LoS} conditions are distinguished: N (\ac{NLoS}), L (\ac{LoS}), and R (regular, i.e., sampled per channel realization based on the scenario's \ac{LoS} probability, 73.8-94.9\%), yielding $4\cdot3=12$ channels in total, denoted as SCENARIO-LoS (e.g. \ac{URB}-N for the urban scenario in \ac{NLoS}). All results use an elevation angle of \SI{70}{\degree}. Experiments repeated at other elevation angles yielded consistent observations. A model trained and evaluated on the same channel is referred to as \emph{matched}, otherwise it is \emph{mismatched}.

Twelve dedicated expert models (experts), one per channel, are trained and denoted by $\mathcal{M}^{\text{SCENARIO-LoS}}$ after their training channel. Each expert has 4 residual blocks and 128 convolutional channels, the baseline architecture used throughout.

All models are trained for up to $3\cdot10^{6}$ iterations with Adam, each step transmitting an \ac{OFDM} grid for a batch of 512 users with \ac{SNR} sampled uniformly from \SI{-15}{dB} to \SI{10}{dB}. The default learning rate is $10^{-3}$, which is reduced to $10^{-4}$ for the $K=12$ \ac{MoE} models (Section~\ref{sec:scenario_aware_arch}). Link-level performance is evaluated over the full training \ac{SNR} range in steps of \SI{0.5}{dB}, with $10^4$ frames and a batch size of 512 per point. For each model, up to 10 independent realizations were trained and evaluated to confirm independence from random initialization and noise. No significant performance differences across realizations were found. Plotted value ranges are individually adapted to highlight the relevant portion of the data. All simulations are executed using the 32 bit floating point (FP32) number format.
\vspace{-10pt}

\subsection{Individual Experts}\label{sec:individual_experts}

\begin{figure}[htbp]
    \centering
\begin{tikzpicture}
\pgfplotsset{compat=1.8}
\definecolor{darkgray176}{RGB}{176,176,176}
\definecolor{darkorange25512714}{RGB}{255,127,14}
\definecolor{forestgreen4416044}{RGB}{44,160,44}
\definecolor{lightgray204}{RGB}{204,204,204}
\definecolor{steelblue31119180}{RGB}{31,119,180}

\definecolor{baseline_cbf}{RGB}{0,77,64}
\definecolor{urb_cbf}{RGB}{216,27,96}
\definecolor{dur_cbf}{RGB}{30,136,229}
\definecolor{sur_cbf}{RGB}{255,193,7}

\pgfplotsset{
compat=1.11,
legend image code/.code={
\draw[mark repeat=2,mark phase=2]
plot coordinates {
(0cm,0cm)
(0.15cm,0cm)        
(0.3cm,0cm)         
};%
}
}

\begin{axis}[
line width = 1pt,
legend cell align={left},
legend style={
    fill opacity=0.8, draw opacity=1, text opacity=1,
    draw=lightgray204, font=\scriptsize,
    at={(0.45,1.26)}, anchor=north,
    legend columns=5,
    /tikz/every even column/.append style={column sep=0.1cm}
},
log basis y={10},
tick align=outside,
tick pos=left,
x grid style={darkgray176},
xlabel={SNR (dB)},
xmajorgrids,
xmin=-9, xmax=0,
xminorgrids,
xtick style={color=black},
y grid style={darkgray176},
ymajorgrids,
ymin=1e-06, ymax=0.5,
yminorgrids,
ymode=log,
height=0.5\linewidth,
width = 0.9\linewidth,
ytick style={color=black},
ytick={1e-06,1e-05,0.0001,0.001,0.01,0.1,1,10},
yticklabels={
  $\displaystyle {10^{-6}}$,
  $\displaystyle {10^{-5}}$,
  $\displaystyle {10^{-4}}$,
  $\displaystyle {10^{-3}}$,
  $\displaystyle {10^{-2}}$,
  $\displaystyle {10^{-1}}$,
  $\displaystyle {10^{0}}$,
  $\displaystyle {10^{1}}$
},
ylabel={BER},
]

\addplot [dur_cbf, mark=none, mark size=2, mark options={solid}]
coordinates {%
(-10.0,2.24845887e-01)
(-9.5,2.03366195e-01)
(-9.0,1.75177044e-01)
(-8.5,1.26281193e-01)
(-8.0,4.17543321e-02)
(-7.5,5.72219978e-03)
(-7.0,2.38340529e-03)
(-6.5,1.38596985e-03)
(-6.0,8.10658270e-04)
(-5.5,4.93854135e-04)
(-5.0,3.01203561e-04)
(-4.5,1.72440639e-04)
(-4.0,1.05916063e-04)
(-3.5,5.63670885e-05)
(-3.0,3.17887931e-05)
(-2.5,1.59519962e-05)
(-2.0,8.96135337e-06)
(-1.5,5.22182450e-06)
(-1.0,2.65693322e-06)
(-0.5,1.49710373e-06)
(0.0,0.93790102e-06)
(0.5,0.05602424e-06)
(1.0,6.66247963e-07)
(1.5,9.12547529e-07)
};
\addlegendentry{$\mathcal{M}^{\text{DUR-R}}$}

\addplot [sur_cbf, mark=x, mark size=2, mark options={solid}]
coordinates {%
(-10.0,2.23327236e-01)
(-9.5,2.01495663e-01)
(-9.0,1.72594979e-01)
(-8.5,1.20352590e-01)
(-8.0,3.69835892e-02)
(-7.5,7.12113397e-03)
(-7.0,3.43854563e-03)
(-6.5,1.94560229e-03)
(-6.0,1.23481662e-03)
(-5.5,7.78916099e-04)
(-5.0,5.02301814e-04)
(-4.5,3.15642328e-04)
(-4.0,1.95179816e-04)
(-3.5,1.23696431e-04)
(-3.0,7.35313193e-05)
(-2.5,4.30397952e-05)
(-2.0,2.55170583e-05)
(-1.5,1.54501099e-05)
(-1.0,9.67131060e-06)
(-0.5,5.61393477e-06)
(0.0,3.89715274e-06)
(0.5,2.32072837e-06)
(1.0,1.82019368e-06)
(1.5,1.09501248e-06)
};
\addlegendentry{$\mathcal{M}^{\text{SUR-R}}$}

\addplot [urb_cbf, mark=x, mark size=2, mark options={solid}]
coordinates {%
(-10.0,2.22838184e-01)
(-9.5,2.00945332e-01)
(-9.0,1.71471451e-01)
(-8.5,1.19174786e-01)
(-8.0,3.45418334e-02)
(-7.5,4.34127258e-03)
(-7.0,1.91305400e-03)
(-6.5,1.16051271e-03)
(-6.0,6.45392704e-04)
(-5.5,3.91663201e-04)
(-5.0,2.38171340e-04)
(-4.5,1.40148230e-04)
(-4.0,7.69323313e-05)
(-3.5,4.21280894e-05)
(-3.0,2.09721958e-05)
(-2.5,1.07743584e-05)
(-2.0,5.85788973e-06)
(-1.5,2.94231226e-06)
(-1.0,1.54277567e-06)
(-0.5,0.80162785e-06)
(0.0,6.96589829e-07)
};
\addlegendentry{$\mathcal{M}^{\text{SUR-N}}$}

\addplot [baseline_cbf, solid, mark=none, mark size=1.5, mark options={solid}]
coordinates {%
(-10.0,2.18673732e-01)
(-9.5,1.96143284e-01)
(-9.0,1.64770081e-01)
(-8.5,1.07735712e-01)
(-8.0,2.35402380e-02)
(-7.5,2.65233041e-03)
(-7.0,1.32859286e-03)
(-6.5,7.40852246e-04)
(-6.0,4.25010397e-04)
(-5.5,2.48063213e-04)
(-5.0,1.26578734e-04)
(-4.5,7.76816481e-05)
(-4.0,3.50285171e-05)
(-3.5,1.97112940e-05)
(-3.0,9.51094344e-06)
(-2.5,5.17647933e-06)
(-2.0,2.98687025e-06)
(-1.5,1.46447243e-06)
(-1.0,8.33976949e-07)
};
\addlegendentry{Perf. CSI}

\addplot [baseline_cbf, dashed, mark=none, mark size=1.5, mark options={solid}]
coordinates {%
(-10.0,4.18041528e-01)
(-9.5,4.09179836e-01)
(-9.0,3.99276156e-01)
(-8.5,3.88685094e-01)
(-8.0,3.76768952e-01)
(-7.5,3.63894591e-01)
(-7.0,3.49798895e-01)
(-6.5,3.34683267e-01)
(-6.0,3.17701848e-01)
(-5.5,2.99756788e-01)
(-5.0,2.79514392e-01)
(-4.5,2.56239158e-01)
(-4.0,2.27803143e-01)
(-3.5,1.80471795e-01)
(-3.0,6.38796340e-02)
(-2.5,6.31817688e-03)
(-2.0,3.25005941e-03)
(-1.5,2.16769992e-03)
(-1.0,1.55520437e-03)
(-0.5,1.12882159e-03)
(0.0,7.79978612e-04)
(0.5,5.61261094e-04)
(1.0,4.21747114e-04)
(1.5,3.01768952e-04)
};
\addlegendentry{Est. CSI}

\end{axis}

\end{tikzpicture}
    \vspace{-5pt}
    \caption{Link-Level performance for different experts and traditional baselines evaluated on channel DUR-R.}
    \label{fig:diff_env_comp}
\end{figure}
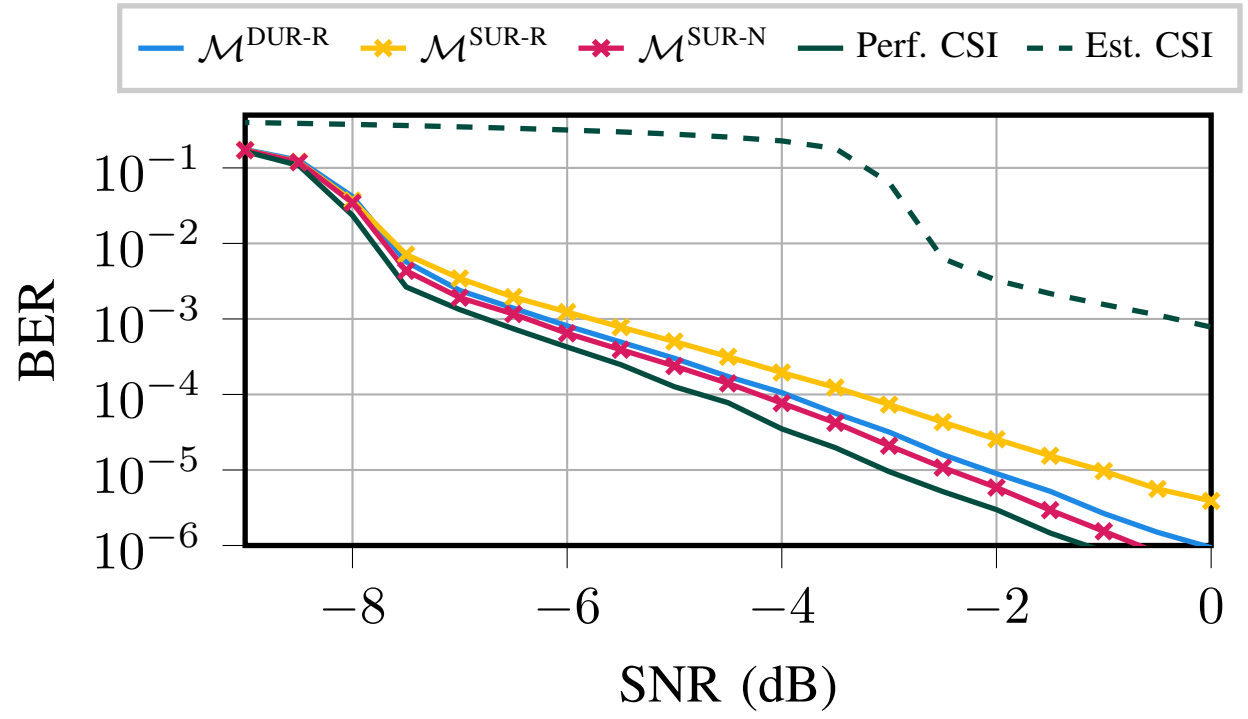

Fig.~\ref{fig:diff_env_comp} shows the performance of a subset of learned receivers on the channel DUR-R, together with conventional receivers with perfect and estimated \ac{CSI} as a baseline. BER decreases with SNR for all receivers and, just like in the terrestrial case, the learned receiver outperforms the estimation based baseline. However, unexpectedly, the receiver trained on DUR-R is not the best. While it outperforms the receiver trained on SUR-R, it is outperformed by SUR-N. This is consistent across 10 independent model realizations and holds true at an elevation angle of \SI{50}{\degree} as well as for minor \ac{OFDM} modifications. This is one of several observed cases in which matching the training and evaluation channels does not yield the best performance.

\begin{figure}
    \centering
    \resizebox{1.0\columnwidth}{!}{\input{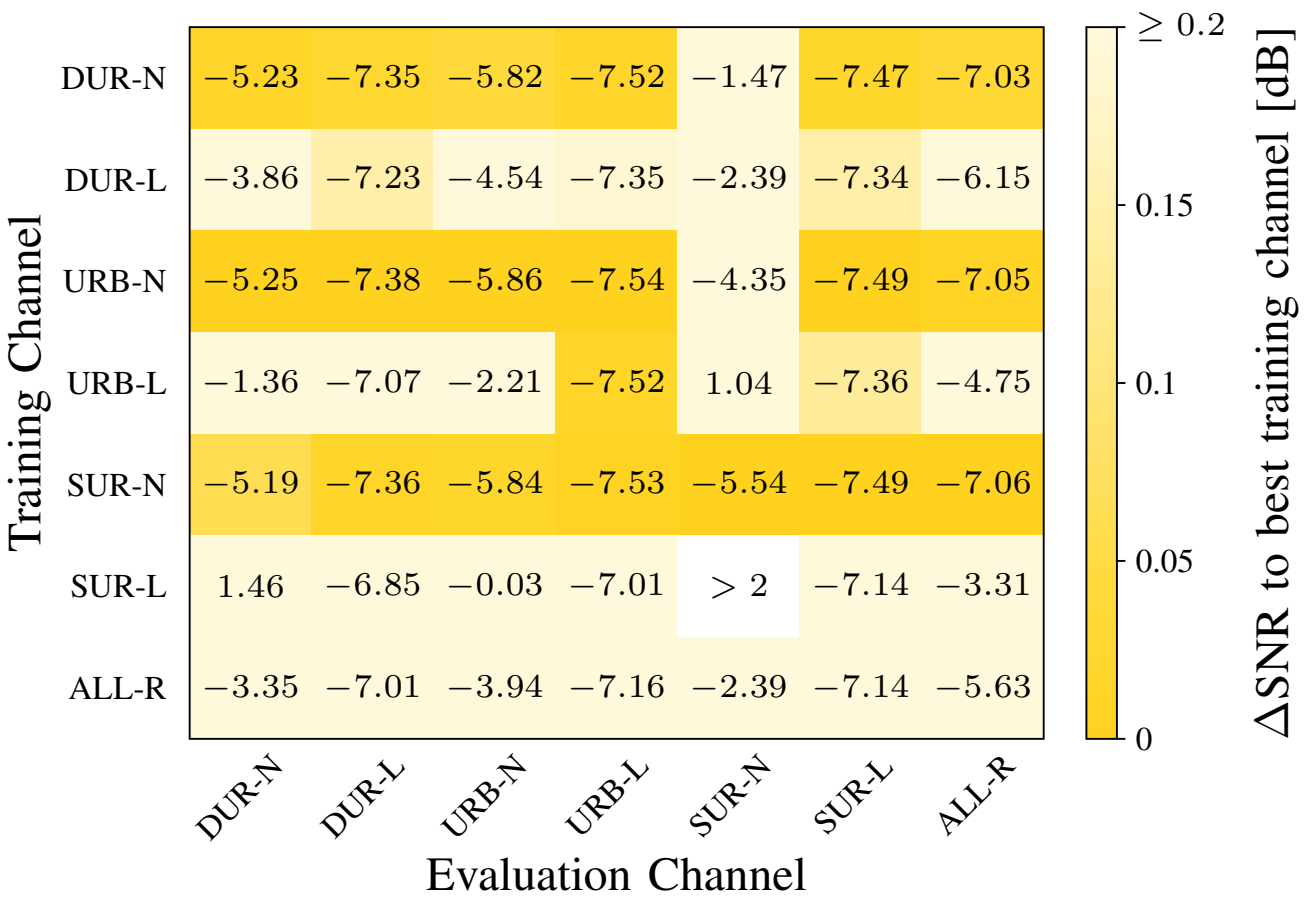}}
    \vspace{-20pt}
    \caption{Confusion matrix showing the SNR [dB] required to reach a BER of $10^{-3}$ for different training channels (rows) and evaluation channels (columns).}
    \label{fig:confusion_matrix}
\end{figure}

To systematically investigate cross-scenario generalization, all 12 experts are evaluated on all 12 channels. The results are summarized in a confusion matrix, using the \ac{SNR} required to achieve a \ac{BER} of $10^{-3}$ as performance metric. For clarity, Fig.~\ref{fig:confusion_matrix} shows only the 6 \ac{LoS}/\ac{NLoS} single scenario channels and the satellite deployment channel ALL-R. A heatmap indicates closeness to the best expert per channel.

Three observations emerge from Fig.~\ref{fig:confusion_matrix}. First, as indicated by the dark colored rows for the \ac{NLoS} experts, training on \ac{NLoS} performs better and generalizes well for most scenarios and towards \ac{LoS}. Second, matching training and evaluation scenario does not always lead to the best performance. While for most channels multiple experts perform very closely and the exact ranking may differ for single realizations, the best experts are always trained on \ac{NLoS}. In particular, ALL-R is significantly outperformed by mismatched experts. Third, SUR-N is an outlier. On most channels all \ac{NLoS} experts perform very closely, SUR-N is the only exception.

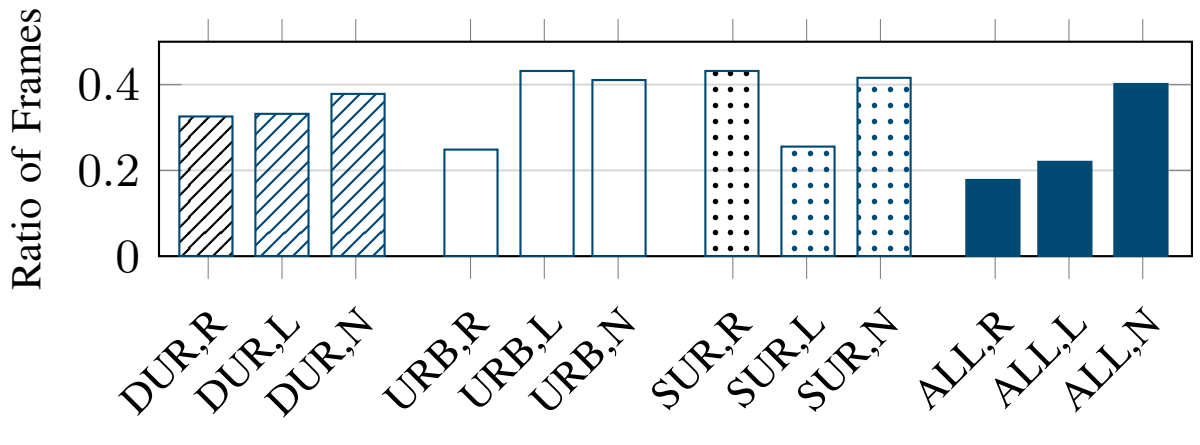
\begin{figure}
    \centering

\begin{tikzpicture}
\begin{axis}[
  ybar,
  width=0.95\linewidth,
  height=3.0cm,
  ymin=-0.0,
  ymax=0.5/1.0,
  xmin=-0.6,
  xmax=12.0,
  xtick={
  0*0.9+0.5*0.0,
  1*0.9+0.5*0.0,
  2*0.9+0.5*0.0,
  3*0.9+0.5*1.0,
  4*0.9+0.5*1.0,
  5*0.9+0.5*1.0,
  6*0.9+0.5*2.0,
  7*0.9+0.5*2.0,
  8*0.9+0.5*2.0,
  9*0.9+0.5*3.0,
  10*0.9+0.5*3.0,
  11*0.9+0.5*3.0},
  xticklabels={
  \footnotesize {DUR,R},
  \footnotesize {DUR,L},
  \footnotesize {DUR,N},
  \footnotesize {URB,R},
  \footnotesize {URB,L},
  \footnotesize {URB,N},
  \footnotesize {SUR,R},
  \footnotesize {SUR,L},
  \footnotesize {SUR,N},
  \footnotesize {ALL,R},
  \footnotesize {ALL,L},
  \footnotesize {ALL,N}
  },
  align={center},
  ylabel={\footnotesize Ratio of Frames},
  bar width=10pt,
  ymajorgrids,
  grid style={gray!30},
  tick label style={font=\small},
  xticklabel style={
    rotate=45,
    anchor=north east
  },
]

\addplot[
  draw=urb_color_dark,
  fill=urb_color_dark,
  pattern=north east lines,
] coordinates {
  (4.25,0.32570/1.0)
};

\addplot[
  draw=urb_color_dark,
  fill=urb_color_dark,
  pattern=north east lines,
  pattern color=urb_color_dark
] coordinates {
  (4.4,0.33187/1.0)
};

\addplot[
  draw=urb_color_dark,
  fill=urb_color_dark,
  pattern=north east lines,
  pattern color=urb_color_dark
] coordinates {
  (4.55,0.37836/1.0)
};

\addplot[
  draw=urb_color_dark,
  fill=urb_color_dark,
  pattern=none,
] coordinates {
  (5.15,0.24852/1.0)
};

\addplot[
  draw=urb_color_dark,
  fill=urb_color_dark,
  pattern=none,
  pattern color=urb_color_dark
] coordinates {
  (5.3,0.43180/1.0)
};

\addplot[
  draw=urb_color_dark,
  fill=urb_color_dark,
  pattern=none,
  pattern color=urb_color_dark
] coordinates {
  (5.4,0.41063/1.0)
};

\addplot[
  draw=urb_color_dark,
  fill=urb_color_dark,
  pattern=dots,
] coordinates {
  (6,0.43180/1.0)
};

\addplot[
  draw=urb_color_dark,
  fill=urb_color_dark,
  pattern=dots,
  pattern color=urb_color_dark
] coordinates {
  (6.15,0.25547/1.0)
};

\addplot[
  draw=urb_color_dark,
  fill=urb_color_dark,
  pattern=dots,
  pattern color=urb_color_dark
] coordinates {
  (6.3,0.41586/1.0)
};

\addplot[
  draw=urb_color_dark,
  fill=urb_color_dark
] coordinates {
  (6.85,0.17773/1.0)
};

\addplot[
  draw=urb_color_dark,
  fill=urb_color_dark,
] coordinates {
  (6.95,0.22000/1.0)
};

\addplot[
  draw=urb_color_dark,
  fill=urb_color_dark,
] coordinates {
  (7.1,0.40148/1.0)
};

\end{axis}
\end{tikzpicture}
    \vspace{-25pt}
    \caption{Ratio of OFDM frames for which each of the twelve experts achieves the lowest bit error count, evaluated on ALL-R at SNR=\SI{-8}{dB}. Ties are credited to all tied models.}
    \label{fig:selection_histogram}
\end{figure}

In addition to the mean-sense result shown in Fig.~\ref{fig:confusion_matrix}, the frequency with which each expert achieves the best per-frame performance is investigated. For the ALL-R channel at an \ac{SNR} of \SI{-8}{dB}, Fig.~\ref{fig:selection_histogram} shows how often each expert achieves the lowest bit error count, with ties credited to all tied experts. 
While \ac{NLoS} experts still perform best frequently, every expert performs best for a notable subset of frames, motivating the combination of multiple experts. Combining experts also intends to reduce sensitivity to rare outlier realizations specific to individual channels.

\vspace{-10pt}

\subsection{Mixture-of-Experts Architectures}
\label{sec:scenario_aware_arch}

Four \ac{MoE} models are developed: two selective and two aggregating architectures. Using all $K=12$ experts yields $\mathcal{M}^{\text{Sel12}}$ and $\mathcal{M}^{\text{Agg12}}$. Since ALL-R comprises channel realizations from the three base scenarios and Fig.~\ref{fig:confusion_matrix} indicates that models trained under \ac{NLoS} conditions generalize better, two additional models, $\mathcal{M}^{\text{Sel3}}$ and $\mathcal{M}^{\text{Agg3}}$, use only the three \ac{NLoS} experts to reduce resource requirements. All \ac{MoE} models include the SUR-N expert. The selector $\mathcal{S}$ and aggregator $\mathcal{A}$ are deliberately kept small: $\mathcal{S}$ uses 128 convolutional channels and a single dense output layer, whereas $\mathcal{A}$ uses 64 convolutional channels. Table~\ref{tab:model_sizes} summarizes the memory requirements of each model.

To distinguish gains attributable to the \ac{MoE} approach from those resulting solely from increased parameter count, an additional model, $\mathcal{M}^{\text{ALL-R,Large}}$, is trained on ALL-R. With six residual blocks and 170 convolutional channels, this model has approximately $2.2$ times the parameters of a single expert. As discussed in Section~\ref{sec:scaling}, further increases in model size yield substantially diminishing returns.

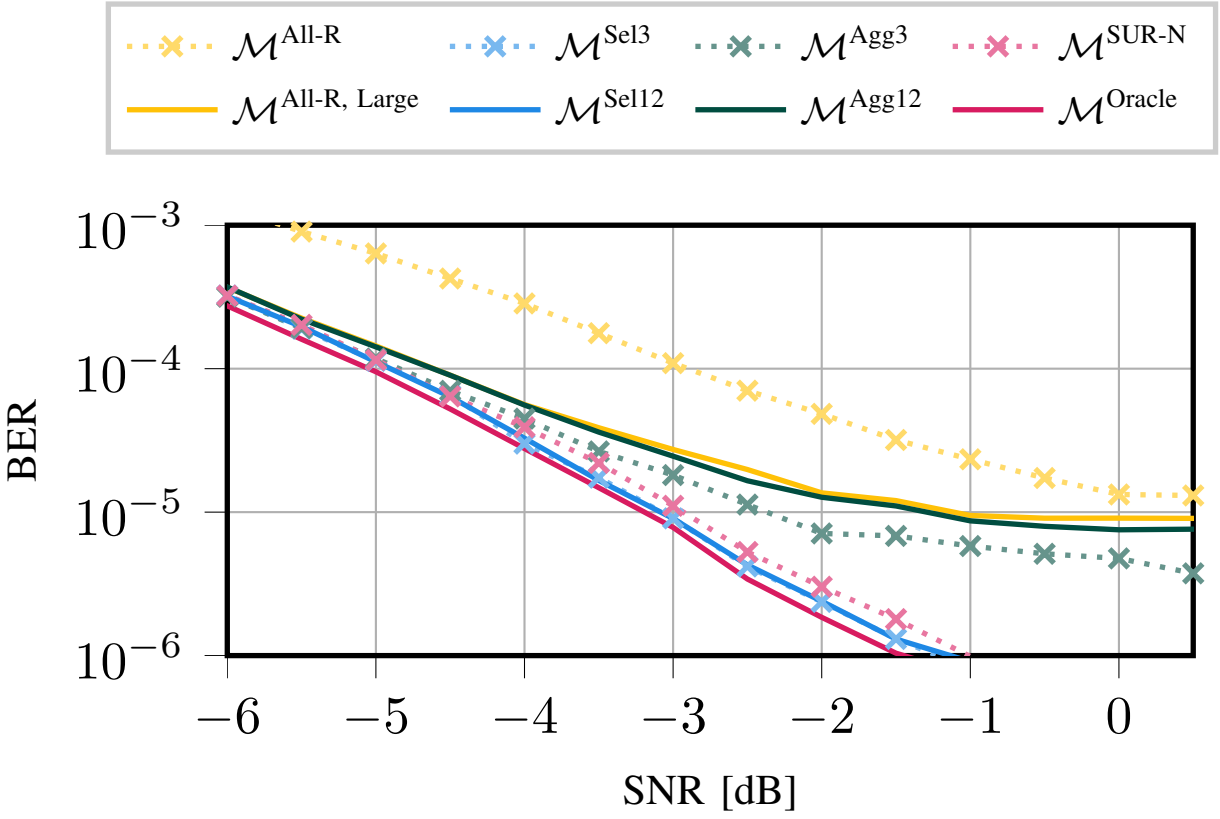
\begin{figure}
    \centering
\begin{tikzpicture}
\pgfplotsset{compat=1.8}
\definecolor{darkgray176}{RGB}{176,176,176}
\definecolor{darkorange25512714}{RGB}{255,127,14}
\definecolor{forestgreen4416044}{RGB}{44,160,44}
\definecolor{lightgray204}{RGB}{204,204,204}
\definecolor{steelblue31119180}{RGB}{31,119,180}

\definecolor{com_cbf}{RGB}{0,77,64}
\definecolor{mis_cbf}{RGB}{216,27,96}
\definecolor{sel_cbf}{RGB}{30,136,229}
\definecolor{all_cbf}{RGB}{255,193,7}

\begin{axis}[
line width = 1pt,
legend cell align={left},
legend style={
    fill opacity=0.8, draw opacity=1, text opacity=1,
    draw=lightgray204, font=\scriptsize,
    at={(0.45,1.52)}, anchor=north,
    legend columns=4,
    /tikz/every even column/.append style={column sep=0.1cm}
},
log basis y={10},
tick align=outside,
tick pos=left,
x grid style={darkgray176},
xlabel={\footnotesize SNR [dB]},
xmajorgrids,
xmin=-6, xmax=0.5,
xminorgrids,
xtick style={color=black},
y grid style={darkgray176},
ymajorgrids,
ymin=1e-06, ymax=1e-03,
yminorgrids,
ymode=log,
height=0.5\linewidth,
width = 0.9\linewidth,
ytick style={color=black},
ytick={1e-06,1e-05,0.0001,0.001,0.01,0.1,1,10},
yticklabels={
  \(\displaystyle {10^{-6}}\),
  \(\displaystyle {10^{-5}}\),
  \(\displaystyle {10^{-4}}\),
  \(\displaystyle {10^{-3}}\),
  \(\displaystyle {10^{-2}}\),
  \(\displaystyle {10^{-1}}\),
  \(\displaystyle {10^{0}}\),
  \(\displaystyle {10^{1}}\)
},
ylabel={\footnotesize BER},
]

\addplot [all_cbf!60!white, mark=x, dotted, mark size=2.5, mark options={solid}]
coordinates {%
(-10.0,2.30915998e-01)
(-9.5,2.10158739e-01)
(-9.0,1.82879917e-01)
(-8.5,1.39647861e-01)
(-8.0,5.71217204e-02)
(-7.5,8.16982007e-03)
(-7.0,3.05675499e-03)
(-6.5,1.87683843e-03)
(-6.0,1.33366860e-03)
(-5.5,9.02746064e-04)
(-5.0,6.35944063e-04)
(-4.5,4.25126743e-04)
(-4.0,2.84967374e-04)
(-3.5,1.76746673e-04)
(-3.0,1.09253456e-04)
(-2.5,7.02399695e-05)
(-2.0,4.84110891e-05)
(-1.5,3.18399372e-05)
(-1.0,2.33575078e-05)
(-0.5,1.72944392e-05)
(0.0,1.32778551e-05)
(0.5,1.30656674e-05)
(1.0,1.20026480e-05)
(1.5,1.11827811e-05)
};
\addlegendentry{$\mathcal{M}^{\text{All-R}}$}

\addplot [sel_cbf!60!white, dotted, mark=x, mark size=2.5, mark options={solid}]
coordinates {%
(-10.0,2.22204618e-01)
(-9.5,1.99516397e-01)
(-9.0,1.69242433e-01)
(-8.5,1.16261970e-01)
(-8.0,3.00762759e-02)
(-7.5,2.37529097e-03)
(-7.0,9.40535474e-04)
(-6.5,5.40188024e-04)
(-6.0,3.23614629e-04)
(-5.5,1.89888022e-04)
(-5.0,1.13336044e-04)
(-4.5,6.54560733e-05)
(-4.0,3.00295593e-05)
(-3.5,1.73926037e-05)
(-3.0,9.05078319e-06)
(-2.5,4.18288147e-06)
(-2.0,2.36853612e-06)
(-1.5,1.30654488e-06)
(-1.0,8.11807737e-07)
};

\addlegendentry{$\mathcal{M}^{\text{Sel3}}$}

\addplot [com_cbf!60!white, dotted, mark=x, mark size=2.5, mark options={solid}]
coordinates {%
(-10.0,2.20980196e-01)
(-9.5,1.98180014e-01)
(-9.0,1.67909591e-01)
(-8.5,1.13775398e-01)
(-8.0,2.81687358e-02)
(-7.5,2.18044453e-03)
(-7.0,8.85312774e-04)
(-6.5,5.13281021e-04)
(-6.0,3.17208838e-04)
(-5.5,1.92136691e-04)
(-5.0,1.17720688e-04)
(-4.5,7.00886017e-05)
(-4.0,4.48827039e-05)
(-3.5,2.65052432e-05)
(-3.0,1.81351356e-05)
(-2.5,1.12897407e-05)
(-2.0,7.10365225e-06)
(-1.5,6.82938103e-06)
(-1.0,5.80477353e-06)
(-0.5,5.12515843e-06)
(0.0,4.75926429e-06)
(0.5,3.74474169e-06)
(1.0,2.89603325e-06)
(1.5,2.40584827e-06)
};
\addlegendentry{$\mathcal{M}^{\text{Agg3}}$}

\addplot [mis_cbf!60!white, dotted, mark=x, mark size=2.5, mark options={solid}]
coordinates {%
(-10.0,2.21877062e-01)
(-9.5,1.99353414e-01)
(-9.0,1.69622746e-01)
(-8.5,1.16345656e-01)
(-8.0,3.04807721e-02)
(-7.5,2.32021744e-03)
(-7.0,8.92279587e-04)
(-6.5,5.34900784e-04)
(-6.0,3.22919142e-04)
(-5.5,2.02070583e-04)
(-5.0,1.15011011e-04)
(-4.5,6.41205195e-05)
(-4.0,3.90922053e-05)
(-3.5,2.20290964e-05)
(-3.0,1.10734019e-05)
(-2.5,5.27566540e-06)
(-2.0,3.02005580e-06)
(-1.5,1.79568679e-06)
(-1.0,9.85645735e-07)
(-0.5,0.00000000e+00)
};
\addlegendentry{$\mathcal{M}^{\text{SUR-N}}$}

\addplot [all_cbf, mark=none, solid, mark size=2.5, mark options={solid}]
coordinates {%
(-15.0,3.58314842e-01)
(-14.5,3.46931751e-01)
(-14.0,3.35417390e-01)
(-13.5,3.23486675e-01)
(-13.0,3.11787970e-01)
(-12.5,2.99231345e-01)
(-12.0,2.85842243e-01)
(-11.5,2.71854233e-01)
(-11.0,2.57159546e-01)
(-10.5,2.40249803e-01)
(-10.0,2.21678659e-01)
(-9.5,1.99270061e-01)
(-9.0,1.69574559e-01)
(-8.5,1.16140248e-01)
(-8.0,3.01932211e-02)
(-7.5,2.36825348e-03)
(-7.0,1.01447818e-03)
(-6.5,6.00754776e-04)
(-6.0,3.68482790e-04)
(-5.5,2.25037166e-04)
(-5.0,1.44140245e-04)
(-4.5,9.01197332e-05)
(-4.0,5.62024833e-05)
(-3.5,3.87382636e-05)
(-3.0,2.73389859e-05)
(-2.5,1.98173739e-05)
(-2.0,1.36238140e-05)
(-1.5,1.20329090e-05)
(-1.0,9.47852025e-06)
(-0.5,9.07490834e-06)
(0.0,9.08238729e-06)
(0.5,9.04679522e-06)
(1.0,9.03271320e-06)
(1.5,9.09708265e-06)
(2.0,9.16423366e-06)
(2.5,7.14087679e-06)
(3.0,9.09067762e-06)
(3.5,7.73045673e-06)
(4.0,9.35095283e-06)
(4.5,7.06760366e-06)
(5.0,7.61842522e-06)
(5.5,6.50405124e-06)
(6.0,6.54252671e-06)
(6.5,6.56057356e-06)
(7.0,7.78936112e-06)
(7.5,5.38104112e-06)
(8.0,6.61303032e-06)
(8.5,6.20153342e-06)
(9.0,7.21276516e-06)
(9.5,6.05457024e-06)
};
\addlegendentry{$\mathcal{M}^{\text{All-R, Large}}$}

\addplot [sel_cbf, mark=none, solid, mark size=2.5, mark options={solid}]
coordinates {%
(-10.0,2.21744518e-01)
(-9.5,1.99458195e-01)
(-9.0,1.69311566e-01)
(-8.5,1.16511228e-01)
(-8.0,3.00852597e-02)
(-7.5,2.33571001e-03)
(-7.0,9.17925677e-04)
(-6.5,5.38594394e-04)
(-6.0,3.19711633e-04)
(-5.5,1.96383940e-04)
(-5.0,1.11554551e-04)
(-4.5,6.37201718e-05)
(-4.0,3.27507539e-05)
(-3.5,1.66446665e-05)
(-3.0,9.03212350e-06)
(-2.5,4.28499127e-06)
(-2.0,2.39403196e-06)
(-1.5,1.30271169e-06)
(-1.0,9.08657255e-07)
(-0.5,8.23560014e-07)
(0.0,5.53014100e-07)
};
\addlegendentry{$\mathcal{M}^{\text{Sel12}}$}

\addplot [com_cbf, solid, mark=none, mark size=2.5, mark options={solid}]
coordinates {%
(-10.0,2.22469572e-01)
(-9.5,2.00324198e-01)
(-9.0,1.70994903e-01)
(-8.5,1.17829429e-01)
(-8.0,3.19557998e-02)
(-7.5,2.66683212e-03)
(-7.0,1.00972714e-03)
(-6.5,6.17320186e-04)
(-6.0,3.70856171e-04)
(-5.5,2.21348394e-04)
(-5.0,1.41897146e-04)
(-4.5,8.99853875e-05)
(-4.0,5.57714283e-05)
(-3.5,3.61316389e-05)
(-3.0,2.45491272e-05)
(-2.5,1.65479174e-05)
(-2.0,1.26987556e-05)
(-1.5,1.10408745e-05)
(-1.0,8.69296578e-06)
(-0.5,7.96552022e-06)
(0.0,7.53439336e-06)
(0.5,7.60788973e-06)
(1.0,7.56999654e-06)
(1.5,7.20873661e-06)
};
\addlegendentry{$\mathcal{M}^{\text{Agg12}}$}

\addplot [mis_cbf, solid, mark=none, mark size=2.5, mark options={solid}]
coordinates {%
(-10.0,2.13398713e-01)
(-9.5,1.90828348e-01)
(-9.0,1.59700147e-01)
(-8.5,1.01361130e-01)
(-8.0,2.05368142e-02)
(-7.5,1.71696373e-03)
(-7.0,8.04108791e-04)
(-6.5,4.66563688e-04)
(-6.0,2.74139484e-04)
(-5.5,1.60350637e-04)
(-5.0,9.52161514e-05)
(-4.5,5.23647097e-05)
(-4.0,2.77523886e-05)
(-3.5,1.47957892e-05)
(-3.0,7.82529614e-06)
(-2.5,3.41702618e-06)
(-2.0,1.84812811e-06)
(-1.5,1.03922931e-06)
(-1.0,7.25470004e-07)
(-0.5,8.31749049e-07)
};

\addlegendentry{$\mathcal{M}^{\text{Oracle}}$}

\end{axis}

\end{tikzpicture}
    \vspace{-5pt}
    \caption{Comparison of the proposed MoE models, the single-architecture baselines, and the oracle, evaluated on ALL-R.}
    \label{fig:multihead_compare}
\end{figure}

Fig.~\ref{fig:multihead_compare} compares all models evaluated on ALL-R.
Training directly on ALL-R performs worst. Increasing model size improves the performance, but still leads to an error floor. 
The model $\mathcal{M}^{\text{SUR-N}}$ performs very well, being only $\approx$\SI{0.5}{dB} off the oracle upper bound.
The selective models perform best, performing in between $\mathcal{M}^{\text{SUR-N}}$ and the oracle with $\approx$\SI{0.25}{dB} distance to both. The performance for both 3 and 12 experts is nearly identical.
The aggregating models underperform despite having access to more computation, which is attributed to $\mathcal{S}$ and $\mathcal{A}$ being deliberately lightweight, limiting their ability to leverage all experts. 

The best deployment choice depends on the constraints. As shown in Table~\ref{tab:model_sizes}, $\mathcal{M}^{\text{Sel3}}$ achieves the best performance at small requirements per inference, while $\mathcal{M}^{\text{SUR-N}}$ achieves similar performance at a smaller stored footprint.

The initial observation was that matching the training and evaluation channels does not always yield the best performance, motivating the investigation of \ac{MoE}-based approaches to address this issue. For the specific deployment channel ALL-R, the expert $\mathcal{M}^{\text{SUR-N}}$ performs well on its own training channel, while also generalizing effectively across all channels. This makes it a particularly strong baseline and limits the potential gains from combining multiple experts. However, such favorable generalization is not expected to hold universally, as other deployment scenarios may involve multiple outlier channels or more pronounced generalization gaps between channels. 

Although the aggregating architectures do not achieve competitive performance at the evaluated model sizes, investigating their trade-offs at larger expert sizes and under more demanding performance requirements remains an objective for future work.

The following hypothesis may explain why $\mathcal{M}^{\text{ALL-R}}$ underperforms even on its own training channel. The \ac{NLoS} case is inherently more challenging, particularly in the \ac{SUR} scenario, which exhibits a distinctive channel characteristic due to its very small delay spread. However, SUR-N realizations account for only approximately $1.7\%$ of ALL-R samples, meaning that their contribution to the training objective is likely negligible. More generally, Fig.~\ref{fig:confusion_matrix} indicates that \ac{NLoS}-trained models generalize well across scenarios, yet \ac{NLoS} conditions are underrepresented in ALL-R. Consequently, the inability to reliably decode SUR-N realizations may contribute to an error floor at high \ac{SNR}.

\begin{table}[tbp]
\centering
\caption{Total \& per inference memory requirements (FP32)}
\label{tab:model_sizes}
\scriptsize
\renewcommand{\arraystretch}{0.9}
\begin{tabular}{lcc}
\hline
Model & Stored & Executed per inference \\
\hline
Single expert & \SI{17.4}{MB} & \SI{17.4}{MB} \\
$\mathcal{M}^{\text{ALL-R,Large}}$ & \SI{37.7}{MB} & \SI{37.7}{MB} \\
$\mathcal{M}^{\text{Sel3}}$ (3 experts and $\mathcal{S}$) & ${\approx}\SI{52.3}{MB}$ & ${\approx}\SI{17.5}{MB}$ \\
$\mathcal{M}^{\text{Agg3}}$ (3 experts and $\mathcal{A}$) & ${\approx}\SI{53.1}{MB}$ & ${\approx}\SI{53.1}{MB}$ \\
$\mathcal{M}^{\text{Sel12}}$ (12 experts and $\mathcal{S}$) & ${\approx}\SI{208.9}{MB}$ & ${\approx}\SI{17.5}{MB}$ \\
$\mathcal{M}^{\text{Agg12}}$ (12 experts and $\mathcal{A}$) & ${\approx}\SI{209.7}{MB}$ & ${\approx}\SI{209.7}{MB}$ \\
Selector $\mathcal{S}$ & ${\sim}\SI{67}{kB}$ & ${\sim}\SI{67}{kB}$ \\
Aggregator $\mathcal{A}$ & ${\sim}\SI{950}{kB}$ & ${\sim}\SI{950}{kB}$ \\
\hline
\end{tabular}
\end{table}

\vspace{-10pt}

\section{Scaling the Model}
\label{sec:scaling}

Having addressed cross-scenario generalization, the focus now shifts to resource constraints. 
Increasing the number of model parameters typically improves performance, but also raises memory, computational, and power requirements. Since the overhead introduced by $\mathcal{S}$ and $\mathcal{A}$ is negligible, this trade-off is investigated at the level of a single expert. Specifically, $\mathcal{M}^{\text{DUR-R}}$ serves as the baseline for architectural scaling, pruning~\cite{pruning}, and \acf{KD}~\cite{hinton2015kd}.

For architectural scaling the baseline architecture contains $N_{\text{Blocks}}=4$ residual blocks and $N_{\text{Conv}}=128$ convolutional channels. $N_{\text{Blocks}}$ is varied as $N_{\text{Blocks}} \in \{2,4,6,8,11\}$ keeping $N_{\text{Conv}}=128$. Conversely, $N_{\text{Conv}}$ is varied over $N_{\text{Conv}} \in \{16,32,48,64,128,256\}$ keeping $N_{\text{Blocks}}=4$.

Pruning is applied as unstructured magnitude pruning~\cite{pruning} with a polynomial sparsity schedule: sparsity increases gradually up to a target $\rho_f \in [0,1]$, interleaved with continued optimization, mask recomputed per weight tensor, avoiding the larger degradation of one-shot pruning at high sparsity.

For \ac{KD}~\cite{hinton2015kd}, the baseline expert is the teacher and a 4-block/64-channel model the student, trained against a frozen teacher using its soft output:
\begin{equation}
\begin{aligned}
&L_{\text{KD}}(\mathbf{C}, \mathbf{Y},N_0) = {}  \gamma \, \mathcal{L}_{\text{SCE}}\Big(\mathbf{C}, \mathcal{M}^{\text{Student}}(\mathbf{Y},N_0)\Big) \\
 &+ (1-\gamma) \, \mathcal{L}_{\text{SCE}}\Big(\sigma\big(\mathcal{M}^{\text{Teacher}}(\mathbf{Y},N_0)\big),\, \mathcal{M}^{\text{Student}}(\mathbf{Y},N_0)\Big),
\end{aligned}
\label{eq:kd_loss}
\end{equation}
where $\gamma$ trades off hard labels against the teacher's soft outputs. The experiments use $\rho_f \in \{0.3, 0.5, 0.8\}$ and $\gamma = 0.7$.

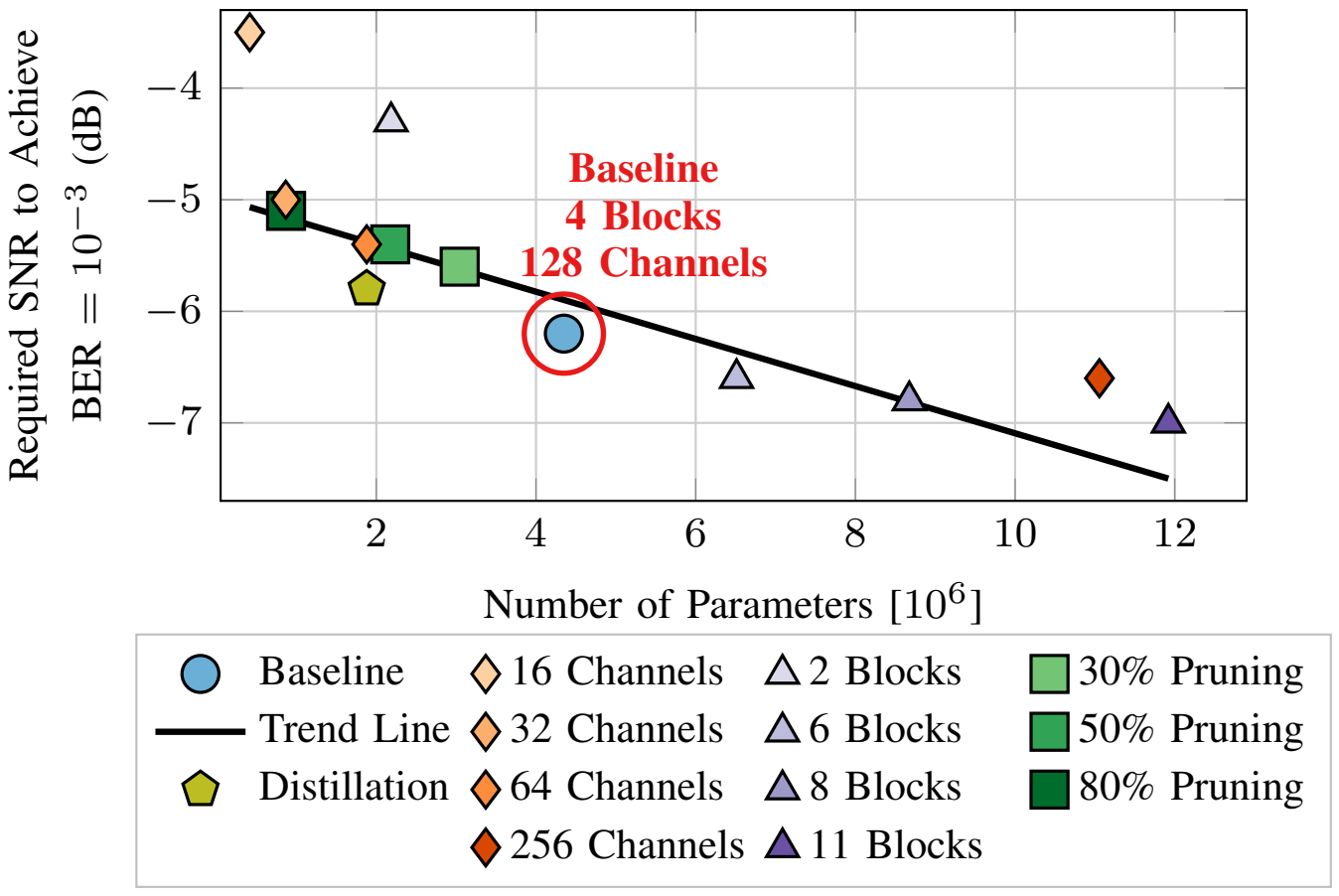
\begin{figure}
    \centering
\begin{tikzpicture}
\definecolor{pareto0}{HTML}{6BAED6}
\definecolor{pareto1}{HTML}{BCBD22}
\definecolor{pareto2}{HTML}{FDD0A2}
\definecolor{pareto3}{HTML}{FDAE6B}
\definecolor{pareto4}{HTML}{FD8D3C}
\definecolor{pareto5}{HTML}{D94801}
\definecolor{pareto6}{HTML}{DADAEB}
\definecolor{pareto7}{HTML}{BCBDDC}
\definecolor{pareto8}{HTML}{9E9AC8}
\definecolor{pareto9}{HTML}{6A51A3}
\definecolor{pareto10}{HTML}{74C476}
\definecolor{pareto11}{HTML}{31A354}
\definecolor{pareto12}{HTML}{006D2C}
\begin{axis}[
  width=0.95\linewidth, height=4.85cm,
  xmin=0.05, xmax=12.9, ymin=-7.699, ymax=-3.300,
  grid=major, grid style={gray!40},
  xlabel={\footnotesize Number of Parameters [$10^6$]},
  ylabel={\footnotesize Required SNR to Achieve\\ \footnotesize BER $=10^{-3}$ (dB)}, 
  ylabel style={align=center},
  mark size=3.5pt,
  tick label style={font=\footnotesize},
  legend columns=4, legend cell align=left,
  legend style={at={(0.5,-0.27)},anchor=north,draw=gray!50,font=\footnotesize,
    /tikz/every even column/.append style={column sep=2pt}},
]
\addplot[only marks,mark=square*,mark options={fill=pareto10,draw=black,line width=0.6pt,solid},forget plot] coordinates {(3.04389,-5.60000)};
\addplot[only marks,mark=square*,mark options={fill=pareto11,draw=black,line width=0.6pt,solid},forget plot] coordinates {(2.17421,-5.40000)};
\addplot[only marks,mark=square*,mark options={fill=pareto12,draw=black,line width=0.6pt,solid},forget plot] coordinates {(0.86968,-5.10000)};
\addplot[black,line width=1.2pt,forget plot] coordinates {(0.41453,-5.06641) (11.91962,-7.49913)};
\addplot[only marks,mark=*,mark options={fill=pareto0,draw=black,line width=0.6pt,solid},forget plot] coordinates {(4.34842,-6.20000)};
\addplot[only marks,mark=pentagon*,mark options={fill=pareto1,draw=black,line width=0.6pt,solid},forget plot] coordinates {(1.87930,-5.80000)};
\addplot[only marks,mark=diamond*,mark options={fill=pareto2,draw=black,line width=0.6pt,solid},forget plot] coordinates {(0.41453,-3.50000)};
\addplot[only marks,mark=diamond*,mark options={fill=pareto3,draw=black,line width=0.6pt,solid},forget plot] coordinates {(0.86592,-5.00000)};
\addplot[only marks,mark=diamond*,mark options={fill=pareto4,draw=black,line width=0.6pt,solid},forget plot] coordinates {(1.87930,-5.40000)};
\addplot[only marks,mark=diamond*,mark options={fill=pareto5,draw=black,line width=0.6pt,solid},forget plot] coordinates {(11.05613,-6.60000)};
\addplot[only marks,mark=triangle*,mark options={fill=pareto6,draw=black,line width=0.6pt,solid},forget plot] coordinates {(2.18522,-4.30000)};
\addplot[only marks,mark=triangle*,mark options={fill=pareto7,draw=black,line width=0.6pt,solid},forget plot] coordinates {(6.51162,-6.60000)};
\addplot[only marks,mark=triangle*,mark options={fill=pareto8,draw=black,line width=0.6pt,solid},forget plot] coordinates {(8.67482,-6.80000)};
\addplot[only marks,mark=triangle*,mark options={fill=pareto9,draw=black,line width=0.6pt,solid},forget plot] coordinates {(11.91962,-7.00000)};

\node[draw=red!90!black!90,circle,minimum size=15pt,inner sep=0pt,line width=1pt] at (axis cs:4.34842,-6.20000) {};
\node[red!90!black!90,align=center,font=\footnotesize\bfseries,anchor=north,inner sep=1pt] at ([yshift=-11pt]axis cs:5.34842,-4.00000) {Baseline\\4 Blocks\\128 Channels};
\addlegendimage{only marks,mark=*,mark options={fill=pareto0,draw=black,line width=0.6pt,solid}}
\addlegendentry{Baseline}
\addlegendimage{only marks,mark=diamond*,mark options={fill=pareto2,draw=black,line width=0.6pt,solid}}
\addlegendentry{16 Channels}
\addlegendimage{only marks,mark=triangle*,mark options={fill=pareto6,draw=black,line width=0.6pt,solid}}
\addlegendentry{2 Blocks}
\addlegendimage{only marks,mark=square*,mark options={fill=pareto10,draw=black,line width=0.6pt,solid}}
\addlegendentry{30\% Pruning}
\addlegendimage{black,line width=1.2pt,mark=none}
\addlegendentry{Trend Line}
\addlegendimage{only marks,mark=diamond*,mark options={fill=pareto3,draw=black,line width=0.6pt,solid}}
\addlegendentry{32 Channels}
\addlegendimage{only marks,mark=triangle*,mark options={fill=pareto7,draw=black,line width=0.6pt,solid}}
\addlegendentry{6 Blocks}
\addlegendimage{only marks,mark=square*,mark options={fill=pareto11,draw=black,line width=0.6pt,solid}}
\addlegendentry{50\% Pruning}
\addlegendimage{only marks,mark=pentagon*,mark options={fill=pareto1,draw=black,line width=0.6pt,solid}}
\addlegendentry{Distillation}
\addlegendimage{only marks,mark=diamond*,mark options={fill=pareto4,draw=black,line width=0.6pt,solid}}
\addlegendentry{64 Channels}
\addlegendimage{only marks,mark=triangle*,mark options={fill=pareto8,draw=black,line width=0.6pt,solid}}
\addlegendentry{8 Blocks}
\addlegendimage{only marks,mark=square*,mark options={fill=pareto12,draw=black,line width=0.6pt,solid}}
\addlegendentry{80\% Pruning}
\addlegendimage{empty legend}\addlegendentry{}
\addlegendimage{only marks,mark=diamond*,mark options={fill=pareto5,draw=black,line width=0.6pt,solid}}
\addlegendentry{256 Channels}
\addlegendimage{only marks,mark=triangle*,mark options={fill=pareto9,draw=black,line width=0.6pt,solid}}
\addlegendentry{11 Blocks}
\addlegendimage{empty legend}\addlegendentry{}
\end{axis}
\end{tikzpicture}
    \vspace{-20pt}
    \caption{Performance-complexity trade-off of scaling, pruning, and distillation, trained and evaluated on DUR-R}
    \label{fig:pareto}
\end{figure}

Fig.~\ref{fig:pareto} summarizes the performance-complexity trade-off, with model size and \ac{SNR} required to achieve an expected BER $=10^{-3}$ on the axes. Both architecture approaches show logarithmically diminishing performance returns at larger parameter counts, with the baseline sitting close to the border of diminishing returns. Reducing $N_{\text{Blocks}}$ degrades the performance faster compared to $N_{\text{Conv}}$, while increasing it reaches the better performance for very large models. Pruning decreases the performance following the trend line. Knowledge distillation improves the performance compared to the regularly trained model of equal size.

The best strategy depends on the target platform. Increasing $N_{\text{Blocks}}$ shows the greatest overall potential, but also increases the number of successive processing steps, which cannot easily be parallelized compared to increasing $N_{\text{Conv}}$. Pruning and knowledge distillation work as intended but require the existence of sparsity compatible hardware or a teacher. Fig.~\ref{fig:pareto} shows a flexible design space, for which the best approach can be selected based on the individual requirements.

\vspace{-10pt}

\section{Synthesis for Satellite Hardware}\label{sec:hardware_deployment}

While Section~\ref{sec:scaling} reported a hardware independent general scaling analysis, this section reports the specific results obtained for the target platform. A single expert is synthesized for the HPDPU-AI satellite processing platform~\cite{dasia}, which executes inference on AMD's Deep Learning Processing Unit (DPU) with a shared \SI{4}{GB} DDR4 memory that necessitates compact models. Because the final deployment on the target hardware is currently ongoing, the reported values are taken from the synthesized model file pre deployment. The model is built from Vitis AI-supported operations using the dedicated tool~\cite{vitis_ai_supported_layers} and compiled into a hardware-optimized \texttt{.xmodel}. 

DPUs operate natively on quantized representations, which influences both memory requirements and potentially performance. Post-training quantization is evaluated in three configurations: 8-bit integer (INT8), 16-bit floating point (FP16), and mixed precision, which uses INT8 for most computations and FP16 for the final, more sensitive layers. At startup, the system calibrates the quantization on a set of input frames. Using up to 20 frames was sufficient to achieve performance equal to the quasi continuous FP32 baseline for all quantizations, at the cost of a slightly longer one-time startup but no effect on steady-state inference speed. Table~\ref{tab:hw_performance} summarizes the resulting metrics relative to FP32.

\begin{table}[t]
\centering
\caption{Deployment metrics of the synthesized single-expert model for different number formats (file size reduction relative to the FP32 model).}
\label{tab:hw_performance}
\footnotesize
\begin{tabular}{lccc}
\hline
Metric & INT8 & Mixed & FP16 \\
\hline
File size reduction & $\sim$75\% & $\sim$70\% & $\sim$50\% \\
Startup setup time & \SI{387}{ms} & \SI{382}{ms} & \SI{450}{ms} \\
Inference latency & \SI{0.01}{ms} & \SI{0.01}{ms} & \SI{0.01}{ms} \\
\hline
\end{tabular}
\end{table}

Quantization reduces model size by approximately 50–75\%, with mixed precision retaining about 80\% of weights in INT8. Calibration adds a small one-time startup cost, while steady-state inference latency remains unchanged across all formats.

Quantizing the parameters offers a significant memory improvement at the cost of only a small startup time for calibration. The best trade-off again depends on the available resources and operation constraints.

\vspace{-10pt}

\section{Conclusion}\label{sec:Conclusion}

This paper investigated neural receivers inspired by the DeepRx architecture for realistic NTN environments, identifying cross-scenario generalization and hardware resource constraints as key deployment challenges. The results revealed that matching training and evaluation channels does not necessarily yield the best performance. To address this mismatch, a \ac{MoE}-based architecture was proposed, in which a selective model combines a small number of scenario-representative experts. It approaches the performance of an oracle upper bound and improves on the best single expert. Furthermore, architectural scaling, knowledge distillation, and pruning were evaluated to characterize the performance-complexity trade-offs of individual experts, providing a flexible design space for resource-constrained satellite platforms. Finally, a receiver was synthesized for an existing satellite ML hardware platform, and its characteristics, including memory footprint under quantization, startup time, and inference latency, were evaluated. These results constitute a step toward deployable learned receivers for NTNs.

\vspace{-5pt}

\bibliographystyle{IEEEtran}
{\scriptsize
\bibliography{main}
}
\end{document}